\documentclass{emulateapj}
\usepackage{amsmath,bm}
\usepackage{natbib}

\begin{document}

\shorttitle{Lensing of 21 cm Power Spectra}
\shortauthors{Mandel \& Zaldarriaga}
\title{Weak Gravitational Lensing of High-Redshift 21 cm Power Spectra}

\author{Kaisey S. Mandel\altaffilmark{1} and Matias Zaldarriaga\altaffilmark{2}}
\affil{Harvard-Smithsonian Center for Astrophysics, 60 Garden St., Cambridge, MA 02138}
\altaffiltext{1}{kmandel@cfa.harvard.edu}
\altaffiltext{2}{Jefferson Laboratory of Physics, Harvard University, Cambridge, MA 02138}

\begin{abstract}
We describe the effects of weak gravitational lensing by cosmological large scale structure on the diffuse emission of 21 centimeter radiation from neutral hydrogen at high redshifts during the era of reionization.  
The ability to observe radial information through the frequency, and thus three-dimensional regions of the background radiation at different redshifts, suggests that 21 cm studies may provide a useful context for studying weak lensing effects.  
We focus on the gravitational lensing effects on both the angular power spectra and the intrinsic, three-dimensional power spectra.
We present a new approach for calculating the weak lensing signature based on integrating differential Fourier-space shells of the deflection field and approximating the magnification matrix.  This reduces the problem of calculating the effect and higher order corrections to solving coupled systems of differential equations.  This method is applied to reionization models of the 21 cm background spectra up to small angular scales over a range in redshift.   
The effect on the angular power spectrum is typically less than one percent on small angular scales, and very small on scales corresponding to the feature imprinted by reionization bubbles, due to the near-scale invariance of the angular power spectrum of the 21 cm signal on these scales. We describe the expected effect of weak lensing on three-dimensional 21 cm power spectra, and show that lensing creates aspherical perturbations to the intrinsic power spectrum which depend on the polar angle of the wavevector.  The effect on the three-dimensional power spectrum is less than a percent on scales $k \lesssim 0.1 \text{ h Mpc}^{-1}$, but can be $\gtrsim 1\%$ for highly inclined modes for $k \gtrsim 1 \text{ h Mpc}^{-1}$.  The angular variation of the lensing effect on these scales is well described by a quartic polynomial in the cosine of the polar angle.
The detection of the gravitational lensing effects on 21 cm power spectra will require very sensitive, high resolution observations by future low-frequency radio arrays.
\end{abstract}


\section{Introduction}

After recombination at the surface of last scattering of the cosmic microwave background (CMB) at $z = 1100$, most of the baryons in the universe are in the form of neutral hydrogen.  Soon after the first galaxies and stars formed, they began to reionize the intergalactic medium.  Understanding the astrophysical processes through which this happened is a major challenge in modern cosmology.  Observations of the Gunn-Peterson trough in high-redshift quasars suggest that the universe was reionized by redshift $z \sim 6$ \citep[e.g.][]{fan02}.  Meanwhile, the large optical depth to Thomson scattering of CMB photons observed by \textit{WMAP} provides an integral constraint on reionization, indicating that reionization began before $z \gtrsim 14$ \citep{spergel03}.  At face value, the evidence suggests that reionization was a long and complicated process; it is a topic of intense current research.

Since neutral hydrogen radiates in the rest frame at a wavelength of 21 cm due to the ``spin-flip'' hyperfine transition, 21 cm tomography has been suggested as a very promising probe of the history of reionization.  This radiation should be observable at earth at low radio frequencies after being cosmologically redshifted.  This 21 cm radiation from the reionization epoch should carry immensely rich information about the reionization process and the formation of structure. Furthermore, since by scanning through frequency one effectively is scanning through different redshifts of 21 cm emission, one has the remarkable ability to probe the evolution of the neutral hydrogen density with time.  Early work studying the potential of 21 cm observations to probe the high-redshift universe was done by, e.g. \citet{scottrees90,madaumeiksinrees97} and \citet{tozzi00}.

Recent research has focused on the expected statistical signal of the 21 cm intensity from the epoch of reionization (EoR) \citep{zfh04, moraleshewitt04, bharadwajali05}.  It was proposed by \citet{zfh04} that one may study the angular power spectra of brightness temperature fluctuations at different frequencies (and cosmic epochs) in order to probe the details of the reionization history.  An alternative approach emphasizes the full three-dimensional nature of the fluctuations \citep{moraleshewitt04}.  Angular observations of the 21 cm sky stepped through different frequencies can be combined to form an image cube by mapping frequency to line-of-sight distance, and such three-dimensional image cubes can be used to probe the statistical properties of the full three-dimensional distribution of HI fluctuations at a particular epoch during reionization.  
The simple three-dimensional structure of the intrinsic fluctuations can be harnessed to remove foregrounds with distinct symmetry properties. 
Data analysis techniques applying this approach have been developed by \citet{moraleshewitt04, morales05, bowmanmoraleshewitt05} and \citet{ moralesbowmanhewitt05}.  Detection of the 21 cm signal from high redshift is among the major goals of planned future low frequency radio arrays, including the Mileura Widefield Array (MWA), the Primeval Structure Telescope (PAST), the Low Frequnecy Array (LOFAR), and the Square Kilometer Array (SKA).

The expected statistical signal detected by future 21 cm observations will be of a complicated nature, resulting from a combination of cosmological sources and astrophysical processes.  The intrinsic 21 cm signal is determined by fluctuations in density, neutral fraction, and the spin temperature describing the relative populations of HI spin states, which is affected by the background radiation field. \citet{fzh04a} and \citet{fzh04b} have shown how inhomogenous reionization will imprint a shoulder feature on the power spectra; hence the power spectra can be an effective discriminant between possible reionization scenarios. \citet{bharadwajali04} have pointed out the importance of the effect of peculiar velocities, which introduce redshift space distortions and are caused by the infall towards (or outflow from) density perturbations.  \citet{barkanaloeb05} have described how these peculiar velocity fluctuations lead to anisotropies in the intrinsic three-dimensional power spectrum $P(\bm{k})$ with characteristic angular dependences on the orientation of $\bm{k}$.  These anisotropies may be exploited to separate out the effects of the density fluctuations and astrophysical processes.  The effects of radiative processes through the spin temperature have recently been considered by \citet{pritchardfurlanetto05} and \citet{hirata05}.

These sources of fluctuations together determine the intrinsic power spectrum, and hence the autocorrelation function,  at any given epoch (or redshift).  As the 21 cm photons emitted by HI stream through the universe, their paths are perturbed stochastically by density inhomogeneities, and their observed directions in the sky are altered.  Hence, gravitational lensing by large scale structure presents another source of fluctuations and modifies the intrinsic power spectra and correlation functions.  These weak lensing modifications are best thought of as secondary fluctuations; they are determined by density fluctuations along the photon path between emission and observation.  Furthermore, they act upon existing fluctuations: without intrinsic brightness fluctuations characterized by the physics of the EoR, lensing would have no effect.

The effect of weak gravitational lensing on diffuse backgrounds has been studied extensively in the context of the CMB.  The effects on the temperature and polarization power spectra have been discussed in \citet{seljak96,zaldarriagaseljak98,hu00,cooray04} and \citet{challinorlewis05}.  Furthermore, the methods of using weak lensing to reconstruct intervening mass distributions has been studied in, e.g. \citet{zaldarriagaseljak99,hu2001a,hu2001b} and \citet{hirataseljak03}.  Future observations of 21 cm backgrounds will present a new source of diffuse emission that will be subject to weak lensing by large scale structure.   In contrast to the CMB, which presents a single, narrow surface at one redshift, the 21 cm emission from different redshifts provides multiple source planes that may be observed in order to study lensing effects.  Indeed, the three-dimensional nature of 21 cm tomography provides a new and interesting context for lensing sudies.  Some initial steps have been taken in this direction.  \citet{pen04} has suggested a method to extract information on the projected lensing convergence from future 21 cm surveys. 

In this paper we focus on the expected weak gravitational lensing effects on the angular and three-dimensional power spectra of 21 cm emission from the era of reionization.  In \S \ref{diffeqsection} we motivate and derive a new differential method for calculating the lensing modifications.  Additionally the lensing effects in certain limiting cases are explored.  In \S \ref{clresults}, we apply these methods to models of the 21 cm angular power spectra. In \S \ref{3dlensingsection}, we discuss the lensing distortions to the three-dimensional power spectra of 21 cm fluctuations, and show that they produce asymmetric perturbations.  These effects have not been considered in previous work.  In \S \ref{detectsection} we compare the magnitude of these effects to expected sensitivites of a future SKA experiment.
Finally, we conclude in \S \ref{conclusion}.

\section{Differential Equation Approach to Weak Lensing}
\label{diffeqsection}

The case of weak gravitational lensing effects on the CMB power spectrum has been thoroughly explored \citep[e.g.][]{seljak96,zaldarriagaseljak98, hu00, challinorlewis05, cooray04}.  The 21 centimeter background from the EoR presents a distinct case in which to study the weak lensing effect.  
The two cases differ significantly in the distribution of fluctuation power on small scales versus large scales.   In the case of the CMB, the intrinsic power is dominated by the acoustic peaks on scales of order $l \sim 100$, while small scale power $l \gtrsim 1000$ is suppressed due to Silk damping.  The contributions to the temperatures fluctuations are confined to angular scales larger than arcseconds.  In the case of the 21 cm background fluctuations from EoR, the power spectrum is dominated by small scale power on scales $l \gtrsim 1000$.  In order to fully understand the effects of weak lensing by large scale structure, one must consider many more decades of scale, and much smaller scales, than in the case of the CMB.

This simple observation has implications for how one calculates lensing effects on the 21 cm background.  Existing methods for calculating the lens effects on power spectra of diffuse backgrounds are similar in that they are perturbative in the deflection angle field, which maps the source plane into the observed image plane.  For example, in the method of Hu (2000), and as extended to higher order in Cooray (2004), the lensed temperature field is expanded in the deflection angle as a Taylor series.  
\begin{equation}\label{texpansion}
\begin{split}
\tilde{T}(\bm{\theta}) &= T(\bm{\theta} + \bm{\delta\theta}(\bm{\theta}))\\ 
&\approx T(\bm{\theta}) + \nabla_i T(\bm{\theta}) \delta\theta^i + \frac{1}{2} \nabla_a \nabla_b T(\bm{\theta}) \delta\theta^a \delta\theta^b + \ldots
\end{split}
\end{equation}
The deflection angles are coupled to higher and higher derivatives of the temperature field.  In cases where there is dominant small scale power, and thus significant fluctuations with wavelengths smaller than the typical deflection angle on arcminute scales, one expects that expansions of the temperature field in which only a few terms are retained are insufficient to describe the lensing effect.  A low-order perturbative expansion of the unlensed temperature in the deflection angle is an inaccurate approximation of the lensed temperature field if the temperature is not sufficiently smooth on scales smaller than and of order the typical deflection angle.  In order to characterize the lensing effect, it is apparent that one must carefully take into account higher order terms in these methods.

There is a caveat in this reasoning.  Since ultimately we wish to understand the lensing effect on the power spectrum, the lensing effect comes through only the two-point function of the product of the temperature at one observed point and another point.  One can show that the calculation depends only on the difference between the deflection angles at the two points, rather than the individual magnitudes of each deflection angle.  This is simply a consequence of the fact that an overall, coherent deflection of the temperature pattern on the sky will not change the pattern.   Hence the relevant physical quantity is the variation of the deflection field over the scale of temperature fluctuations.

A series solution for the lensed power spectrum is obtained by forming the lensing correlation function $\langle \tilde{T}(\bm{\theta}) \tilde{T}(\bm{\theta}') \rangle$, and performing the transform analytically.  The result is
\begin{equation}\label{series}
\begin{split}
&\tilde{C}^T_l = C^T_l e^{-l^2 R} + \sum_{n=1}^{\infty} \frac{1}{n!}   \int \frac{d^2 k_1}{(2\pi)^2} \dotsi \int \frac{d^2 k_n}{(2 \pi)^2}\\ 
&\times \left[ \bm{k_1} \cdot (\bm{l} - \bm{k_1} - \ldots - \bm{k_n}) \right]^2 \dotsm 
\left[ \bm{k_n} \cdot (\bm{l} - \bm{k_1} - \ldots - \bm{k_n}) \right]^2 \\ 
&\times C^{\phi}_{k_1} \ldots C^{\phi}_{k_n} C^T_{ | \bm{l} - \bm{k_1} - \ldots - \bm{k_n} | } e^{ -| \bm{l} - \bm{k_1} - \ldots - \bm{k_n} |^2 R }
\end{split}
\end{equation}
where $ R \equiv (1/4\pi) \int d\ln k\, k^4 C^{\phi}_k $ \citep{cooray04}.   In order to compute the result to the $N$th order in $C^\phi_l$, one must expand the exponentials and retain all integrals with up to $N$ factors of the lensing potential power spectrum, $C^{\phi}_k$.

In the harmonic space expansions of \citet{hu00} and \citet{cooray04}, this dependence on only the relative deflection angle manifests itself in cancellations between positive and negative integrals at a given order in the lensing potential power spectrum.  When there is significant small scale power, the individual terms at a given order can become very large, but the total contribution of these terms may be very small due to these cancellations between positive and negative terms.  A correct computation of higher order contributions to the lensed power spectrum must carefully account for delicate cancellations between such terms.  Hence, the presence of significant small-scale power alone does not guarantee that appropriately summed higher-order corrections will be important.  The degree to which such cancellations occur generally depends on the shape of the intrinsic power spectrum.  Nevertheless, one must check these high order contributions in order to determine whether or not the lensing calculation has converged properly on small angular scales.

In the harmonic space expansion, however, the calculation of higher order terms becomes numerically difficult.  Whereas the leading order lensing contribution (at order $C^{\phi}_l$) is a simple integral over the two-dimensional space of the lensing potential wavevector, the second order correction requires the evaluation of a four-dimensional integral.  In general, the computation of the $(C^{\phi\phi}_l)^N$ corrections requires the evaluation of a $2N$-dimensional numerical integral over the unlensed temperature spectrum and $N$ lensing potential power spectra.  In order to compute the lensing effects on the 21 cm spectrum, in which there is significant small scale power in several decades, these integrations must be done over many decades of scale.  At each order, one must correctly handle the delicate cancellations between these multi-dimensional integrals over decades in scale.  This procedure for accounting for higher-order effects becomes numerically exhaustive and slow, and may lead to inaccurate estimates.

The other approach used in calculating the lensed CMB power spectrum is described in \citet{seljak96} and \citet{zaldarriagaseljak98}.  In this formulation, the two-point autocorrelation function of the lensed temperature field is computed numerically, and the transform of this result is computed to calculate the power spectrum.  This is also essentially an expansion in the full deflection angle coupled to derivatives of the temperature field.  Indeed, one can show that these two approaches are equivalent when one does not terminate either series or expands each to the same order in the deflection dispersion.  A downside to this method is the need to transform back and forth between position and Fourier space.  This involves numerical integrations with Bessel functions which become highly oscillatory at small scales.  Furthermore, the transforms obscure the effects of the lensing power spectrum on modifying the temperature power spectrum.  A purely harmonic calculation has the appeal of transparency, as all relevant quantities interact in the Fourier domain.

The difficulties apparent in these existing methods motivate us to develop a new method to handle the lensing calculation in the case of the 21 cm background.  In this section we present a new method using a differential equation approach.  This method avoids the pitfalls of the above methods.  First, the differential approach allows a straightforward computation of higher order corrections without requiring evaluation of irreducible high-dimensional integrals.  The differential approach allows us at once to sum up many terms from all orders in the series.   Cancellations between terms, corresponding to differences in deflections, are carefully done shell by shell in Fourier space as a function of a single wavenumber cutoff $\Lambda$.  Extensions of the calculation to include finer approximations are straightforward to derive.  The systems of differential equations are solved entirely in harmonic space, avoiding highly oscillatory Bessel transforms back and forth between position and harmonic spaces.  Furthermore, the harmonic space solution allows for illuminating analytic simplifications in certain limits.  Our new approach is easy to implement, requiring only one-dimensional integration and a differential equation solver.

\subsection{Formalism}

Our analysis  will be carried out in the formalism of the flat-sky approximation, which is sufficient on small angular scales and small sections of the sky.  Since we will apply this method to the 21 cm emission, where most of the power is at small scales, this approximation should be sufficient for our purposes.  We expect that corrections due to the curvature of the sky will be small in the regime studied in this work.  In general, for high-precision and low multipoles, one should take into account curvature effects (e.g. \citet{hu00}, \citet{challinorlewis05}).  In the flat sky limit, an observeable $X$ on the sky can be expanded in Fourier modes rather than spherical harmonics, e.g. 
\begin{equation}
X(\bm{\theta}) = \int \frac{d^2 l}{(2\pi)^2} X(\bm{l}) e^{i\bm{l} \cdot \bm{\theta}}
\end{equation}
with power spectra
\begin{equation}
\langle X(\bm{l}) X(\bm{l'}) \rangle = (2\pi)^2 \delta^2 (\bm{l}-\bm{l'}) C^{XX}_l.
\end{equation}

The lensing modes are usually approximated as Gaussian random variables.  The fluctuations of lens quantities, e.g. the projected potential, deflection angle and effective convergence, are described by the power spectra, which are integrals over the matter fluctuations along the line of sight.  The convergence power spectrum to a source at comoving radial distance $r_s$ and redshift $z_s$ in the small scale limit using the Limber approximation is \citep[e.g.][]{bartelmannschneider01}:
\begin{equation}
C^{\kappa\kappa}_l = \frac{9 H_0^4 \Omega_M^2}{4 c^4} \int^{r_s} dr \frac{W(r, r_s)^2}{a^2(r)} P_{\delta}\left(k = \frac{l}{\chi(r)}, r \right)
\end{equation}
where $W(r, r_s) = \chi(r_s -r)/\chi(r_s)$, $P_{\delta}$ denotes the matter power spectrum, and $\chi$ is the comoving angular diameter distance.  The power spectrum of the projected lensing potential is related simply through the Laplace equation, $2\kappa(\bm{\theta}) = \nabla^2 \phi(\bm{\theta})$, so that $l^4 C^{\phi\phi}_l =  4 C^{\kappa\kappa}_l$.

\begin{figure}[t]
\centering
\includegraphics[angle=90,scale=0.4]{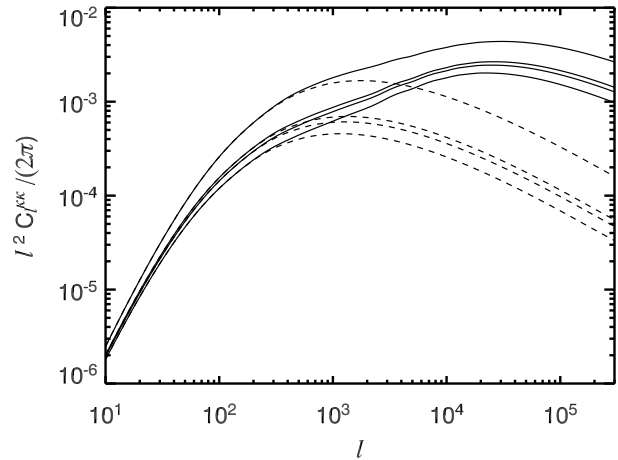}
\caption{\label{convergenceplot}Angular power spectrum of the lensing convergence.  The solid curves are the non-linear spectra for $z = 1100$, $15$, $12$, and $8$.  The dashed curves are the corresponding linear spectra.}
\end{figure}

We use the code CMBFAST \citep{cmbfast} to generate the transfer functions and to compute the lensing convergence power spectrum.  On small scales it is important to account for the non-linear evolution of the matter power spectrum.  We use the method of \citet{peacockdodds96} to calculate the non-linear approximation.  The linear and non-linear power spectra for several redshifts in the range of interest are plotted in Figure \ref{convergenceplot}.   These will generally depend on the cosmological model.  For the rest of this paper, we adopt a $\Lambda$CDM model with these values of the cosmological parameters: $\Omega_M = 0.3$, $\Omega_b = 0.046$, $\Omega_\Lambda = 0.7$, $h= 0.70$, with mass fluctuations on scales of $8 \text{ h Mpc}^{-1}$, $\sigma_8 = 0.88$, in accordance with the latest data \citep{spergel03}.

\subsection{Derivation}

In this section, we derive a new method for calculating the lensing effect on the power spectrum.  Our strategy differs fundamentally from previous approaches.  Instead of finding a solution perturbative in the full deflection angle, we expand in a differential shell of modes of the deflection field.  This leads us to a differential equation solution for the lensing effect in Fourier space.  We show how this strategy leads to a hierarchy of systems of differential equations.  A method for truncating the hierachy in the limit of weak lensing is proposed, leading to simple closed systems as first and second approximations.  We explore various limits of these equations to gain intuition of the lensing process in Fourier space.

In what follows, we will adopt the following conventions.  The unobserved, source-plane quantities are described by non-tilde notation, e.g. the unlensed, intrinsic temperature power spectrum is  $C^T_l$ and the source temperature is $T(\bm{\beta})$.  Quantities that are affected through lensing are described by tilde notation.  Hence, the lensed temperature power spectrum is $\tilde{C}^T_l$ and the lensed, observed temperature is $\tilde{T}(\bm{\theta})$.

The lensed temperature field is related to the unlensed field through the deflection angle.
\begin{equation}
\tilde{T}(\bm{\theta}) = T(\bm{\beta}) = T(\bm{\theta}+\bm{\delta\theta}(\bm{\theta})).
\end{equation}
The deflection field can be decomposed into Fourier components.
\begin{equation}
\bm{\delta\theta}(\bm{\theta}) = \int \frac{d^2k}{(2\pi)^2} i \bm{k} \phi(\bm{k}) e^{i \bm{k} \cdot \bm{\theta}}
\end{equation}
Now we introduce the notion of a harmonic-space ``momentum'' cutoff on lensing quantities.  Define the cutoff-dependent deflection field including lensing modes up to wavenumber $|\bm{k}| \le \Lambda$,
\begin{equation}
\bm{D}_{\bm{\theta}}(\Lambda) = \int^{\Lambda}_0 \frac{d^2k}{(2\pi)^2} i \bm{k} \phi(\bm{k}) e^{i \bm{k} \cdot \bm{\theta}},
\end{equation}
and the cutoff-dependent ``source position'' $\bm{\beta}_\Lambda(\bm{\theta}) = \bm{\theta} + \bm{D}_{\bm{\theta}}(\Lambda)$.
The temperature field lensed by lensing modes with wavenumber up to $|\bm{k}| \le \Lambda + \delta\Lambda $ is
\begin{equation}
\begin{split}
\tilde{T}_{\Lambda+\delta\Lambda} (\bm{\theta}) &= T(\bm{\theta} + \bm{D}_{\bm{\theta}}(\Lambda + \delta\Lambda)) \\
&=T(\bm{\beta}_{\Lambda} + \bm{d}_{\bm{\theta}})
\end{split}
\end{equation}
where we have defined the differential deflection element including only lensing modes with $\Lambda \le |\bm{k}| \le \Lambda+\delta\Lambda$:
\begin{equation}
\bm{d}_{\bm{\theta}} \equiv \left[ \bm{D}_{\bm{\theta}}(\Lambda+\delta\Lambda) -\bm{D}_{\bm{\theta}}(\Lambda) \right] 
\end{equation}
One may interpret the cutoff $\Lambda$ as a low-pass filter on lensing quantities that allows one to calculate the lensing effects of only long-wavelength modes on the lensed quantities.  In any numerical calculation, there is a practical cutoff on the maximum frequency of a Fourier mode included in the calculation of lensed quantities.  We have simply introduced this cutoff explicitly, and use it as a tool to regulate the calculation of the lensed quantities.  We will show how the lensing problem reduces to a system of differential equations in the cutoff, and this system will be evolved to a high enough $\Lambda$ so that the calculation in the regime of interest has converged and thus is independent of higher frequency spatial modes of the lens.

The lensed temperature correlation function, lensed by deflection modes $k < \Lambda$ is formed as:
\begin{equation}
\begin{split}
&\tilde{C}^T_{\Lambda+\delta\Lambda}(|\bm{\Delta\theta}|) = \langle \tilde{T}_{\Lambda+\delta\Lambda} (\bm{\theta}) \tilde{T}_{\Lambda+\delta\Lambda} (\bm{\theta'}) \rangle \\
&= \langle T(\bm{\beta_\Lambda} + \bm{d}) T(\bm{\beta_\Lambda'} + \bm{d}) \rangle = \langle T(\bm{\beta_\Lambda}) T( \bm{\beta'} + \Delta \bm{d}) \rangle \\
&= \langle \tilde{T}_\Lambda(\bm{\theta}) \left[ T(\bm{\beta_\Lambda'}) + \ T_n(\bm{\beta_\Lambda'}) \Delta d^n + \frac{1}{2} T_{ia}(\bm{\beta_\Lambda'}) \Delta d^a \Delta d^i \right] \rangle \\
&= \langle \tilde{T}_\Lambda \tilde{T}_\Lambda' \rangle - \frac{1}{2} \langle \frac{\partial}{\partial {\beta_\Lambda^i}} T(\bm{\beta}_\Lambda)  \frac{\partial}{\partial {\beta_\Lambda^a}'} T(\bm{\beta}_{\Lambda}') \rangle \langle \Delta d^a \Delta d^i  \rangle
\end{split}
\end{equation}
where primed and unprimed quantities correspond to the coordinates $\bm{\theta}$ and $\bm{\theta'}$, respectively, and $\bm{\Delta\theta} = \bm{\theta}-\bm{\theta'}$ is the displacement vector.  Gradients of the temperature field with respect to the source position $\bm{\beta}_\Lambda$ are denoted by subscripts, e.g. $\partial T/\partial\beta_\Lambda^i = T_i(\bm{\beta_\Lambda})$.
We have expanded to second-order in the relative differential deflection angle $\Delta \bm{d} \equiv \bm{d}' - \bm{d}$ and used the translational invariance property of the average. The Einstein summation convention for repeated indices is implied in each term, and $i, j = 1, 2 = \theta_x, \theta_y$.
Here we have expressed the temperature lensed by lensing modes $|\bm{k}| \le \Lambda + \delta\Lambda$ in relation to the lensed temperature just including lensing modes $|\bm{k}| \le \Lambda$, both at the observed position, $\bm{\theta}$.  In the last two terms, the first ensemble averages are over the temperature and lensing modes up to $|\bm{k}| \le \Lambda$, and the second averages are over the lensing modes in the Fourier shell $ \Lambda \le |\bm{k}| \le \Lambda + \delta\Lambda$, and we have assumed the lensing mode amplitudes are distributed as Gaussian random variables. 
This arrangement of terms emphasizes that the lensing effect depends only on differences between deflection angles at different points, not on their absolute values.  This is simply a consequence of the fact that the correlation functions only depend upon differences in position.  Since we will divide through by $\delta\Lambda$ and take its limit to zero, the neglected terms in the Taylor expansion involving higher order products of the relative deflection will vanish; hence, this expansion becomes exact.

In order to evaluate the derivatives on the source plane with respect to $\bm{\beta}_\Lambda$ we must make a change of coordinates to relate them to the observed coordinates $\bm{\theta}$:
\begin{equation}\label{coordchange}
\frac{\partial}{\partial\beta_{\Lambda}^i} = \frac{\partial\theta^j}{\partial\beta_{\Lambda}^i} \frac{\partial}{\partial\theta^j} \equiv M^j_{i}(\Lambda) \frac{\partial}{\partial\theta^j} 
\end{equation}
The matrix $M^j_i(\Lambda) \equiv \partial\theta^j / \partial\beta_{\Lambda}^i $ is the inverse of the magnification matrix of the $\Lambda$-filtered lensing quantities $A^a_b(\Lambda) = \partial\beta_{\Lambda}^a / \partial\theta^b$: 
\begin{equation}
 M^j_i A^i_b = \frac{\partial\theta^j}{\partial\beta_{\Lambda}^i} \frac{\partial\beta_{\Lambda}^i}{\partial\theta^b} = \frac{\partial\theta^j}{\partial\theta^b} = \delta^j_b.
\end{equation}
The magnification matrix is explicitly:
\begin{equation}
\mathbf{A} = \begin{pmatrix} 1 + \phi_{,11}& \phi_{,12} \\ \phi_{,12} & 1 + \phi_{,22} \end{pmatrix} = \begin{pmatrix} 1 + \kappa + \gamma_1 & \gamma_2 \\ \gamma_2 & 1 + \kappa - \gamma_1 \\ \end{pmatrix}
\end{equation} 
where we have defined the usual convergence and shears, $\kappa \equiv \frac{1}{2} (\phi_{,11} + \phi_{,22})$, $\gamma_1 \equiv \frac{1}{2} (\phi_{,11} - \phi_{,22})$, and $\gamma_2 \equiv \phi_{,12}$, and all quantities are implicitly cut off at lensing modes $|\bm{k}| = \Lambda$.  The inverse is then simply,
\begin{equation}\label{generalm}
\begin{split}
\mathbf{M} &= \frac{1}{(1+\kappa)^2 - (\gamma_1^2 + \gamma_2^2)}   \begin{pmatrix} 1 + \kappa - \gamma_1 & -\gamma_2 \\ -\gamma_2 & 1 + \kappa + \gamma_1 \end{pmatrix} \\
&\approx \mathbf{I} - \begin{pmatrix} \kappa + \gamma_1 & \gamma_2 \\ \gamma_2 & \kappa - \gamma_1 \end{pmatrix}.
\end{split}
\end{equation}

With these substitutions, the equation for the correlation function expressed in observed coordinates is simply
\begin{equation}\label{correlation2}
\begin{split}
&\tilde{C}^T_{\Lambda+\delta\Lambda}(\Delta\theta) - \tilde{C}^T_{\Lambda}(\Delta\theta)\\ 
&=  -\frac{1}{2} \langle  \left[ {M^j_i} \partial_j \right] \tilde{T}_{\Lambda} \left[  {M^b_a}' \partial_b' \right] \tilde{T}_{\Lambda}' \rangle \langle \Delta d^i \Delta d^a \rangle 
\end{split}
\end{equation}
This equation relates the lensed correlation function with lensing cutoff $\Lambda + \delta\Lambda$ to the lensed correlation function with cutoff $\Lambda$; i.e. it is the change from integrating over a differential shell of lensing modes.  The correlations of the differential deflections will be proportional to $\delta\Lambda$ in the limit $\delta\Lambda \rightarrow 0$.  We will see that this leads to a differential equation for the lensed correlation function as a function of the cutoff $\Lambda$, or for the lensed power spectrum when we transform directly to harmonic space.

There is an alternative way to derive the above results that motivates another intuitive picture.  The full deflection angle tells us how to map the source plane temperature to the image plane temperature.  One can imagine intermediate steps in this mapping, which are image planes that are lensed through the cutoff-truncated deflection field, $\bm{D}_{\bm{\theta}} (\Lambda)$, each labelled by the cutoff wavenumber, $\Lambda$.  Just as the lens equation relates the intensities between the source plane and the lensed image plane, one can derive a relation between these intermediate lensed image planes.  Consider the image lensed by modes up to $\Lambda + \delta\Lambda$, $\tilde{T}_{\Lambda + \delta\Lambda} (\bm{\theta})$.  One can relate this to a previously lensed image plane, $\tilde{T}_{\Lambda}(\bm{\theta})$, lensed only by lensing modes up to $\Lambda$, through the lens equation.  Let the vector $\bm{X}_{\bm{\theta}}$ be the effective infinitesimal deflection angle that maps the image at $\Lambda$ to the image at $\Lambda + \delta\Lambda$.
\begin{equation}
\tilde{T}_{\Lambda+\delta\Lambda} (\bm{\theta}) = \tilde{T}_\Lambda(\bm{\theta} + \bm{X}_{\bm{\theta}})
\end{equation}
The effective deflection $\bm{X}_{\bm{\theta}}$ can be determined by mapping each side to the source plane:
\begin{equation}
\tilde{T}_{\Lambda + \delta\Lambda} (\bm{\theta}) = T(\bm{\theta} + \bm{D}_{\bm{\theta}}(\Lambda) + \bm{d}_{\bm{\theta}})
\end{equation}
\begin{equation}
\tilde{T}_\Lambda (\bm{\theta} + \bm{X}_{\bm{\theta}}) = T(\bm{\theta} +\bm{X_\bm{\theta}}+ \bm{D}_{\bm{\theta} + \bm{X}}(\Lambda))
\end{equation}
Since the two expressions are equal,  $\bm{X}_{\bm{\theta}} + \bm{D}_{\bm{\theta} + \bm{X}}(\Lambda) = \bm{D}_{\bm{\theta}}(\Lambda) + \bm{d}_{\bm{\theta}}$.  Next the deflection field is expanded in the infinitesimal vector $\bm{X}$: $D^a_{\bm{\theta}+\bm{X}}(\Lambda) - D^a_{\bm{\theta}} = (\partial_b D^a_{\bm{\theta}}) X^b$.  Then we have
\begin{equation}
d^a_{\bm{\theta}} = (\delta^a_b + \partial_b D^a_{\bm{\theta}}(\Lambda) ) X^b= A^a_b(\Lambda) X^b 
\end{equation}
so that $X^a_{\bm{\theta}} = M^a_b d^b_{\bm{\theta}}$.  Thus the effective incremental deflection between two images labelled by their lensing cutoffs, $\Lambda + \delta\Lambda$ and $\Lambda$, is simply the incremental deflection $\bm{d}_{\bm{\theta}}$ rotated and magnified by the inverse magnification matrix of the lensing modes up to $\Lambda$.  This relation can then be used to derive the above Eq. \eqref{correlation2}, by expanding the temperatures at each position, $\tilde{T}_{\Lambda + \delta\Lambda}(\bm{\theta})$, in $\bm{X}$ to relate it to the previous lensed image, $\tilde{T}_{\Lambda}(\bm{\theta})$.

\subsubsection{The $\mathbf{M = I}$ approximation}

As a first approximation, if the lens is sufficiently weak, we may approximate the inverse magnification matrix as identity $M^b_a(\Lambda) \simeq \delta^b_a$.  In doing so, terms proportional to $\kappa$, $\gamma_1$, and $\gamma_2$ from Eq. \eqref{generalm} are neglected.  This is justified in the weak lensing regime in which the elements of the distortion tensor are much smaller than unity.  Note that this approximation depends only on the strength of the lensing field and not the underlying temperature power spectrum that is being lensed.  We can see from Figure \ref{convergenceplot} that the variance in the convergence is much less than unity.  In this approximation, the lensing equation then simplifies to:
\begin{equation}
\tilde{C}^T_{\Lambda+\delta\Lambda}(\Delta\theta) - \tilde{C}^T_{\Lambda}(\Delta\theta) = -\frac{1}{2} \langle \partial_i \tilde{T}_\Lambda \partial_a' \tilde{T}_\Lambda' \rangle \langle \Delta d^i \Delta {d^a}' \rangle 
\end{equation}
where $\partial_i \equiv \partial/\partial\theta^i$ denotes derivatives with respect to the observed coordinate.
This is evaluated by decomposing each quantity into Fourier components and averaging the coefficients as $\langle \tilde{T}_{\Lambda}(\bm{l})^* \tilde{T}_{\Lambda}(\bm{l'}) \rangle = (2\pi)^2 \tilde{C}^T_l(\Lambda) \delta^2(\bm{l}-\bm{l'})$ and $\langle  \phi(\bm{k})^* \phi(\bm{k'}) \rangle = (2\pi)^2 C^{\phi}_k \delta^2(\bm{k}-\bm{k'})$.  The lensed power spectrum is obtained by applying a Fourier transform ($\int d^2 \Delta\theta\, e^{-\bm{l'} \cdot (\bm{\theta} -\bm{\theta'})}$ ) 
\begin{equation}
\begin{split}
\Delta \tilde{C}^T_l (\Lambda) = \frac{\delta\Lambda \Lambda C^{\phi}_{\Lambda}}{(2\pi)^2} \Bigl( &\int d\phi_\Lambda \left[ \bm{\Lambda} \cdot (\bm{l} - \bm{\Lambda} )\right]^2 \tilde{C}^T_{|\bm{l}-\bm{\Lambda}|} (\Lambda) \\
&-  \left[\bm{\Lambda} \cdot \bm{l} \right]^2 \tilde{C}^T_{l} (\Lambda) \Bigr) \\
\end{split}
\end{equation}
and taking the limit $\delta\Lambda \rightarrow 0$.  In the limit, this simplifies to an ordinary differential equation for evolving the temperature power spectrum with the cutoff $\Lambda$.  
\begin{equation}\label{approxde}
\begin{split}
\frac{d \tilde{C}^T_l\left(\Lambda\right)}{d\ln \Lambda} = -\frac{l^2 \Lambda^4 C_\Lambda^\phi}{2\pi}&\int_0^{2\pi}\frac{d\phi_\Lambda}{2\pi} \Bigl[ \tilde{C}_l^T\left(\Lambda\right) \cos^2(\phi_\Lambda) \\
&- \Bigl(\frac{\Lambda}{l}-\cos(\phi_\Lambda)\Bigr)^2 \tilde{C}_{|\bm{l}-\bm{\Lambda}|}^T(\Lambda)\Bigr] 
\end{split}
\end{equation}
where $|\bm{l} - \bm{\Lambda}|^2 = l^2 + \Lambda^2 -2 l \Lambda \cos(\phi_\Lambda)$.
This solution contains contributions from all orders in $C^{\phi}_k$, and the error contribution begins at order $O((C^{\phi}_k)^2)$ in the sense of the series solution, Eq. \eqref{series}.
This approximate differential equation is simple to solve, and should be adequate in the limit of very weak lensing, when the magnification matrix does not deviate significantly from identity. 

There are two limiting cases of interest.  For $\Lambda \ll l$, and sufficiently smooth power spectra, we may expand the right-hand side in $x = \Lambda/l$ and perform the integral.  In this limit, the two terms in the integral cancel, and the residual terms are suppressed by a factor of $x^2$:
\begin{equation}\label{lowlambda}
\begin{split}
&\frac{d \tilde{C}^T_l\left(\Lambda\right)}{d\ln \Lambda} = \frac{l^2 \Lambda^4 C_\Lambda^\phi}{4\pi} \Bigl[0 x^0  \\
&+ \Bigl( 2 \tilde{C}_l^T\left(\Lambda\right) + \frac{7}{4} \frac{d\tilde{C}_l^T\left(\Lambda\right)}{d\ln l} + \frac{3}{8} \frac{d^2 \tilde{C}_l^T\left(\Lambda\right)}{d\ln l^2}\Bigr)x^2 + O(x^4) \Bigr] \\
&= \frac{ \Lambda^6 C_\Lambda^\phi}{4\pi} \left[2 \tilde{C}_l^T\left(\Lambda\right) + \frac{7}{4} \frac{d\tilde{C}_l^T\left(\Lambda\right)}{d\ln l} + \frac{3}{8} \frac{d^2 \tilde{C}_l^T\left(\Lambda\right)}{d\ln l^2}  + O(x^2) \right]
\end{split}
\end{equation}
We see that in this limit, in which the differential shell of lensing modes has a wavelength much longer than that of the temperature fluctuation mode, the leading order lensing effect, proportional to $x^0$ in the first line, vanishes.  Physically, this corresponds to a picture in which the deflection field is coherent over the scale of the temperature fluctuations.  Temperature fluctuations on small scales are gravitationally deflected together, resulting in an overall shift that does not change the fluctuation power on those scales. In this limiting case, the vanishing of the leading term is independent of the power in the lensing mode.  Thus, even though the RMS amplitude of the incremental deflection angle may be large compared to the fluctuation with wavenumber $l$, in this particular limit the leading lensing effect vanishes.  
The remaining fractional lensing effect is proportional to $\Lambda^6 C^{\phi}_\Lambda /(8\pi) =  \Lambda^2 C^{\kappa}_\Lambda /(2\pi)$, which is the power per logarithmic interval in the convergence.  As one can see from Figure \ref{convergenceplot}, this is a small dimensionless fraction, which declines steeply on large scales as $\Lambda \rightarrow 0$.
This intuition demonstrates that low $\Lambda$ lensing modes are relatively unimportant, unless $l \sim \Lambda$. The lensing behavior then will generally depend on the power of lensing fluctuations at that scale and the shape of the temperature spectrum. 

It has been pointed out by \citet{hu00} that, since the power in the deflection field $\Lambda^4 C^{\phi}_\Lambda$ peaks at a low multipole, $\Lambda \sim 40$ in $\Lambda$CDM, the lensing effect on even high-$l$ temperature modes must arise from large-scale power in the deflections.  This reasoning, however, fails to fully account for the cancellations among deflection angles over the scale of the temperature fluctuations.
We have shown, in Eq. \eqref{lowlambda}, that in this limit, $\Lambda \ll l$, the relevant quantity determining the differential lensing effect, after accounting for coherent shifts, is actually the power per log interval of the convergence, $\Lambda^6 C^{\phi}_\Lambda /(8\pi)$, which is quite small on large scales and peaks at much smaller scales (Fig. \ref{convergenceplot}).  We will see in \S \ref{clresults} that, in the case of 21 cm fluctuations, the contribution by large-scale lens modes ($\Lambda \lesssim 50$) to the total lensing effect on smaller scale temperature fluctuations is negligible.

In the opposite limit, $\Lambda \gg l$, one may expand the right-hand side in $y = l/\Lambda$.
The result in terms of the variance per logarithmic interval, $\Delta^{2T}_l \equiv l^2 C^T_l /(2\pi)$, is
\begin{equation}
\begin{split}
&\frac{d \Delta^{2T}_l(\Lambda)}{d \ln \Lambda} = \frac{\Lambda^6 C^{\phi}_\Lambda}{2\pi} y^2 \Bigl[ \Delta^{2T}_\Lambda(\Lambda) - \frac{1}{2}\Delta^{2T}_l(\Lambda) \\ 
&+ y^2 \left(-\frac{1}{2}\Delta^{2T}_\Lambda(\Lambda) + \left. \frac{1}{4} \frac{d^2 \Delta^{2T}_k(\Lambda)}{d \ln k^2} \right\vert_{k=\Lambda} \right) +O(y^4) \Bigr]
\end{split}
\end{equation}
In this limit, there is a small wavelength deflection mode lensing a long wavelength temperature mode.  Power on the scale $l$ is modified, and the magnitude of the effect is determined by a combination of the convergence power at the scale $\Lambda$ and the temperature spectrum.  The overall effect is determined by the convergence power, $\Lambda^6 C^{\phi}_\Lambda \sim \Lambda^2 C^{\kappa}_\Lambda$, suppressed by a factor $y^2$; hence we expect that the lensing effect will shut off in the limit, so long as the convergence power is slowly varying on sufficiently small scales.  From Figure \ref{convergenceplot} we see that, whereas the convergence power is increasing on large scales $\Lambda \ll 10^3$, it is fairly flat or decreasing on small scales $\Lambda \gtrsim 10^3$.
The leading order effect is determined by the distribution of temperature fluctuations on large scales ($l$) versus small scales ($\Lambda$).  If the fluctuations are dominated by small scale power, $\Delta^{2T}_\Lambda \gg \Delta^{2T}_l$, the small scale lensing modes produce power on scale $l$.  In this picture, a short wavelength lensing mode can interact with a temperature mode of the same wavelength to produce long wavelength power.  If there is negligible small scale power, $\Delta^{2T}_\Lambda \ll \Delta^{2T}_l$, small scale lensing modes remove power in fluctuations at scale $l$.  This is intuitively the case, for example, for the damping tail of the CMB, in which lensing deflections create small scale fluctuations against a long wavelength gradient \citep{zaldarriaga00}. In this case, there is little intrinsic power in small scale temperature modes $l \gtrsim 4000$, and small scale ($\Lambda$) deflections of large-scale gradients produce small scale temperature fluctuations on scales $l \sim \Lambda$.  Consequently, since lensing preserves total variance, power is removed from the acoustic regime $l \lesssim 2000$.  The above limit demonstrates this overall shifting of fluctuation power.

In these limits we were able to replace an integral over $\tilde{C}^T_l\left(\Lambda\right)$ with its local derivatives.  Hence these formulae will be useful in a numerical implementation for sufficiently smooth power spectra.   Furthermore, these local derivatives of the temperature power spectrum with respect to $l$ emphasize the importance of the shape of the power spectrum in determining the lensing effect.

We note that in cases in which Eq. \eqref{approxde} is a sufficent approximation, and the lensing modifications to the power spectrum are small for all $\Lambda$, one may make the replacement $\tilde{C}^T_l(\Lambda) \approx C^T_l$ on the right-hand side of \eqref{approxde}.  Then the right-hand side can be immediately integrated, and the solution is
\begin{equation}\label{husoln}
\begin{split}
\tilde{C}^T_l(\Lambda) - C^T_l = \int_0^\Lambda &\frac{d^2 k}{(2\pi)^2} C^{\phi}_k \Bigl\{ -(\bm{k} \cdot \bm{l} )^2 C^T_l \\
&+ C^T_{|\bm{l}-\bm{k}|} [\bm{k} \cdot (\bm{l} - \bm{k})]^2 \Bigr\}
\end{split}
\end{equation}
which, as $\Lambda \rightarrow \infty$, is the first-order harmonic-space solution of \citet{hu00}, or, equivalently, the first order terms of the full series solution, Eq. \eqref{series}, obtained by the simplest expansion of the temperature field to second order in (full) deflection angles.  Hence, we see that the solution to Eq. \eqref{approxde} includes this most basic approximation, but sums up higher order contributions by having the current lensed power spectrum $\tilde{C}^T_l(\Lambda)$ on the right hand side of Eq. \eqref{approxde}.

The difference between the lensing solutions, Eq. \eqref{husoln} and Eq. \eqref{approxde} is in the nature of the approximations.  In Eq. \eqref{husoln} the terms that are dropped to make the approximation are those corresponding to the higher order terms in the Taylor expansion, Eq. \eqref{texpansion}, of the form $\frac{1}{n!} \partial_{i_1} \ldots \partial_{i_N} T(\bm{\theta}) \delta\theta^{i_1} \ldots \delta\theta^{i_N}$.  In Eq. \eqref{approxde}, the terms that are dropped are proportional to the elements of the distortion tensor, and are of order $O(\kappa(\Lambda))$ smaller compared to the terms retained, i.e. those corresponding to the identity part of the inverse magnification matrix.  It is in this sense that Eq. \eqref{approxde} sums the series Eq. \eqref{series} in a different way than Eq. \eqref{husoln}.

\subsubsection{Poisson Power Spectrum}

The solution to Eq. \eqref{approxde} is simple in the special case of a Poisson power spectrum. In the Poisson case, $C^T_l = $ constant is independent of $l$.  Lensing acts to remap uncorrelated points, hence the lensed power spectrum remains Poisson.  One can then take $\tilde{C}^T_l(\Lambda)$ as independent of $l$ and remove it from the integral in Eq. \eqref{approxde}.  The solution to the equation is then:
\begin{equation} 
\tilde{C}^T_l = C^T_l \exp(\int \frac{d^2 k}{(2\pi)^2} 4 C^{\kappa}_k) \approx C^T_l ( 1+ 4\langle \kappa^2 \rangle)
\end{equation}
where $C^{\kappa}_k$ is the power spectrum of the convergence.  Thus the temperature spectrum is modified by multiplication by a constant factor determined by the total power in the convergence.   However, the spectrum remains Poisson, i.e. independent of $l$, under the action of lensing.  

We can understand this by considering a simple model.  Consider a region with temperature fluctuations, described by a Poisson spectrum up to some minimum scale $\sim k_{max}^{-1}$, magnified by a constant convergence so that $\tilde{T}(\bm{\theta}) = T(\mu^{-1} \bm{\theta})$, where $\mu$ is the magnification factor.  Then the lensed correlation function is 
\begin{equation}
\begin{split}
\langle \tilde{T}(\bm{\theta}^1) \tilde{T}(\bm{\theta}^2) \rangle &= \langle T(\mu^{-1}\bm{\theta}^1) T(\mu^{-1}\bm{\theta}^2) \rangle \\ 
&= \int^{k_{max}} \frac{d^2 k}{(2\pi)^2}  C^T_k e^{i\mu^{-1}\bm{k} \cdot \bm{\Delta\theta}} \\ 
&= \int^{k_{\max}/\mu} \frac{d^2 l}{(2\pi)^2} \mu^{2} C^T_{l\mu} e^{i\bm{l}\cdot\bm{\Delta\theta}}.
\end{split}
\end{equation}  
In the case of a Poisson spectrum, then $\tilde{C}^T_l = \text{const} \times C^T_l$ on scales $l < k_{max}/\mu$.  One can see that the small scale cutoff $k_{max}$ is rescaled in order to preserve the total variance.  

\subsubsection{Scale-Invariant Power Spectrum}

Suppose at some cutoff $\Lambda$, the variance $\Delta^{2T}_l (\Lambda) \equiv l^2 \tilde{C}^T_l (\Lambda)/(2\pi)$ is approximately constant independent of wavelength, at least within the range of integration.  Equation \eqref{approxde} simplifies as:
\begin{equation}
\frac{d\Delta^{2T}_l(\Lambda)}{d \ln \Lambda} = \frac{ \Lambda^6 C^{\phi}_\Lambda}{4\pi} \Delta^{2T}_l (\Lambda) \begin{cases} 0 & \text{if $\Lambda \le l$}, \\ \frac{l^2}{\Lambda^2}\left[1-( \frac{l}{\Lambda})^2 \right] & \text{if $\Lambda > l$}. \end{cases}
\end{equation}
This shows that small scale temperature modes are not affected by lensing by larger scale lensing modes at scales $\Lambda < l$, independently of $l$ in this regime.  Furthermore, much shorter wavelength lensing modes $\Lambda \gg l$ also have an asymptotically negligible lensing effect on  mode $l$.  If the convergence power $\Lambda^6 C^{\phi}_\Lambda \sim \Lambda^2 C^{\kappa}_\Lambda /(2\pi)$ is slowly varying on small scales $\Lambda \gtrsim l$, then a particular temperature mode $l$ is only affected by lensing modes around a narrow range around $\Lambda \sim \sqrt{2} l$.  As can be seen in Figure \ref{convergenceplot}, the convergence power is indeed slowly varying on angular scales $l \gtrsim 1000$.  In the approximation that the lensing modification of the temperature power spectrum is small, the fractional lensing effect is roughly $\Delta C^T_l/C^T_l \approx \Delta^{2\kappa}_s / 2$, where $\Delta^{2\kappa}_s = s^2 C^{\kappa}_s / (2\pi)$ is the convergence variance per logarithmic interval at scale $s = l\sqrt{2}$.

The limit of the long wavelength lensing mode can be understood physically.  Consider a slowly varying convergence lensing a region that is approximately scale-invariant on smaller scales.  The lensing will result in slowly varying changes in scale over the region.  However, due to the small scale invariance, the appearance of the region is unchanged.

If the lensing mode has a wavelength somewhat smaller than that of a particular long wavelength temperature mode of the scale invariant region, the deflection field will interact with the fluctuation on that scale and modify the power.  The modification on that scale will be proportional to the convergence power spectrum at that scale.

\subsection{General Approach}
\label{generalapproach}

The lensing equation derived above, Eq. \eqref{approxde}, may be considered a first approximation to the full solution.
In this section we derive the general series of differential equations.  We have seen above that the equation governing the change in the temperature correlation function is:
\begin{equation}\label{topeq}
\begin{split}
\Delta\tilde{C}^T_{\Lambda}(\Delta\theta) &\equiv \tilde{C}^T_{\Lambda + \delta \Lambda}(\Delta \theta) - \tilde{C}^T_{\Lambda} (\Delta\theta) \\
&= -\frac{1}{2} \langle T_i(\bm{\beta}_\Lambda) T_j({\bm{\beta}_\Lambda}') \rangle \langle \Delta d^i \Delta d^j \rangle
\end{split}
\end{equation}
where $\bm{\beta}_\Lambda \equiv \bm{\beta}_\Lambda(\bm{\theta}) $ is the source position corresponding to the observed position $\bm{\theta}$, including lensing modes up to wavenumber $\Lambda$, ${\bm{\beta}_\Lambda}'$ is defined analagously. For notational convenience and clarity, derivatives of the source temperature with respect to the source position $\bm{\beta_\Lambda}$ are denoted by subscripts, $T_i(\bm{\beta}_\Lambda)$, and derivatives of the observed, lensed temperature field with respect to the observed coordinate, $\bm{\theta}$, are denoted as $\partial_i \tilde{T}$.  

In the simplest approximation, we evaluated the right hand side of the equation by approximating the slopes of the temperature field on the unobserved source plane by the observed slope on the lensed image plane.
\begin{equation}
\begin{split}
T_i(\bm{\beta}_\Lambda(\bm{\theta})) &\equiv \frac{\partial}{\partial \beta^i_\Lambda} T(\bm{\beta}_\Lambda(\theta))\\  
&= M^a_i(\bm{\theta},\Lambda) \frac{\partial}{\partial \theta^a} \tilde{T}_{\Lambda}(\bm{\theta}) \approx \partial_i \tilde{T}(\bm{\theta})
\end{split}
\end{equation}
where we have projected to the observed image plane and introduced the inverse magnification matrix $M^a_b(\bm{\theta},\Lambda) \equiv \partial\theta^a/\partial\beta_\Lambda^b$, which is then approximated as identity, generating a relative error of leading order  $O(\kappa)$.

In order to make a better approximation, we must determine an equation to describe the correlator of first derivatives on the right hand side of Eq. \eqref{topeq}.
\begin{equation}
\begin{split}
&\langle T_i(\bm{\beta}_{\Lambda + \delta\Lambda}) T_j(\bm{\beta}_{\Lambda + \delta\Lambda}') \rangle = \langle T_i(\bm{\beta}_\Lambda + \bm{d}) T_j(\bm{\beta}_\Lambda' + \bm{d}') \rangle \\ &= \langle T_i(\bm{\beta}_\Lambda) T_j(\bm{\beta}_\Lambda' + \Delta \bm{d}) \rangle \\
&= \langle T_i(\bm{\beta}_\Lambda) \left[ T_j(\bm{\beta}_\Lambda') + \frac{1}{2} T_{jab}(\bm{\beta}_\Lambda')  \langle \Delta d^a \Delta d^b \rangle \right] \rangle \\
&= \langle T_i(\bm{\beta}_\Lambda) T_j(\bm{\beta}_\Lambda') \rangle - \frac{1}{2} \langle T_{ia}(\bm{\beta}_\Lambda) T_{jb}(\bm{\beta}_\Lambda') \rangle \langle \Delta d^a \Delta d^b \rangle
\end{split}
\end{equation}
The change in the first derivative correlator is determined by the second derivatives on the source plane.  A similar derivation can be done for the second derivative correlator.  We get a hierarchy of coupled equations:
\begin{align}\label{3coupled}
\Delta \tilde{C}^T(\Delta\theta) &= -\frac{1}{2} \langle T_i T_j' \rangle \langle \Delta d^i \Delta d^j \rangle \\
\label{3coupled2}\Delta \langle T_i T_j' \rangle &= - \frac{1}{2} \langle T_{ia} T_{jb}' \rangle \langle \Delta d^a \Delta d^b \rangle \\
\label{3coupled3}\Delta \langle T_{ia} T_{jb}' \rangle &= - \frac{1}{2} \langle T_{iac} T_{jbe}' \rangle \langle \Delta d^c \Delta d^e \rangle
\end{align}
and so on. The system of equations is generally not closed, as the change in the correlator of $n$th temperature derivatives is determined by the correlator of $(n+1)$th derivatives.  However, at any level of the hierarchy, one may terminate and close the system by approximating derivatives on the source plane with derivatives on the image plane in an analogous manner as in the first approximation, making an error $O(\kappa)$ smaller than the terms retained at that level.  For example, we may close the above system at the third equation with the approximation, $\langle T_{iac} T_{jbe}' \rangle \approx \partial_c \partial_e' \langle T_{ia} T_{jb}' \rangle$.

\subsection{Spin Decomposition}

In this section we illustrate how one can take the position space equations from the previous section and turn them into coupled differential equations for power spectra in Fourier space.  The correlators of temperature derivatives can be broken down simply in Fourier space into power spectra of $E$ and $B$ type spin quantities.  This approach is analogous to the spin decomposition of polarization.  First, we illustrate  how to decompose the first derivative correlators and derive Fourier space equations.

Define the vector $\bm{v}$ as the gradient of the source temperature on the source plane as a function of the observed position, $\bm{\theta}$, implicitly through its dependence on the source position $\bm{\beta}_\Lambda(\bm{\theta})$, which is also dependent upon the current cutoff scale $\Lambda$.
\begin{equation}
v_i(\bm{\theta}) \equiv T_i(\bm{\beta}_\Lambda(\bm{\theta}))
\end{equation}
From this the complex spin 1 and spin -1 quantities can be defined
\begin{equation}
\begin{split}
\tilde{v}(\bm{\theta}) \equiv (v_x + iv_y)(\bm{\theta}) &= \int \frac{d^2 l}{(2 \pi)^2} a_1(\bm{l}) e^{i\phi_l} e^{i\bm{l} \cdot \bm{\theta}}\\ \tilde{v}(\bm{\theta})^* \equiv (v_x - iv_y)(\bm{\theta}) &= \int \frac{d^2 l}{(2 \pi)^2} a_{-1}(\bm{l}) e^{-i \phi_l} e^{i\bm{l} \cdot \bm{\theta}}.
\end{split}
\end{equation}
Reality requires $a_{-1}(\bm{l}) = -a_1^*(-\bm{l})$. If we define
\begin{equation}
\begin{split}
E(\bm{l}) &= \frac{1}{2} (a_1(\bm{l}) + a_{-1}(\bm{l})) \\  B(\bm{l}) &= \frac{1}{2} i(a_1(\bm{l}) - a_{-1}(\bm{l})) ,
\end{split}
\end{equation}
then these quantities\footnote{These should not be confused with the $E$ and $B$ modes of spin 2 polarization quantities.} can be rewritten:
\begin{equation}
\tilde{v}(\bm{\theta}) \equiv (v_x + iv_y)(\bm{\theta}) = \int \frac{d^2 l}{(2 \pi)^2} \left[ E(\bm{l}) - iB(\bm{l}) \right] e^{i \phi_l} e^{i\bm{l} \cdot \bm{\theta}}
\end{equation}
\begin{equation}
\tilde{v}(\bm{\theta})^* \equiv (v_x - iv_y)(\bm{\theta}) = \int \frac{d^2 l}{(2 \pi)^2} \left[ E(\bm{l}) + iB(\bm{l}) \right] e^{-i \phi_l} e^{i\bm{l} \cdot \bm{\theta}}
\end{equation}
where $E$ and $B$ denote the $E$-type and $B$-type parts of the vector quantity with opposite parity.  This can be seen by defining $E(\bm{\theta}) = \int d^2 l/(2 \pi)^2\, E(\bm{l}) e^{i\bm{l} \cdot\bm{\theta}}$ and $B(\bm{\theta}) =  \int d^2 l/(2 \pi)^2 \, B(\bm{l}) e^{i\bm{l} \cdot\bm{\theta}}$.  These can be written as:
\begin{equation}
2 E(\bm{\theta}) = \int \frac{d^2 l}{(2 \pi)^2} e^{i\bm{l} \cdot \bm{\theta}}  \int d^2 \theta e^{-i\bm{l} \cdot \bm{\theta}} \left[ e^{-i\phi_l} \tilde{v}(\bm{\theta}) + e^{i\phi_l} \tilde{v}^*(\bm{\theta}) \right]
\end{equation}
\begin{equation}
-2i B(\bm{\theta}) =  \int \frac{d^2 l}{(2 \pi)^2} e^{i\bm{l} \cdot \bm{\theta}}  \int d^2 \theta e^{-i\bm{l} \cdot \bm{\theta}} \left[ e^{-i\phi_l} \tilde{v}(\bm{\theta}) - e^{i\phi_l} \tilde{v}^*(\bm{\theta}) \right]
\end{equation}
Under a reflection across the $x$-axis, $\phi_l \rightarrow -\phi_l$, $v_y \rightarrow -v_y$, $\tilde{v} \rightarrow \tilde{v}^*$, $\tilde{v}^* \rightarrow \tilde{v}$.  Thus we see that $E$ is even under parity, $E \rightarrow E$, and $B$ is odd under parity, $B \rightarrow -B$. A similar argument may be made for reflections across the $y$-axis.
Translational and parity invariance dictate:
\begin{equation}
\langle E(\bm{l}) E^*(\bm{l}') \rangle = (2 \pi)^2 C^{EE}_l \delta^2(\bm{l} - \bm{l}')
\end{equation} 
\begin{equation}
\langle B(\bm{l}) B^*(\bm{l}') \rangle = (2 \pi)^2 C^{BB}_l \delta^2(\bm{l} - \bm{l}') 
\end{equation} 
\begin{equation}
\langle E^*(\bm{l}) B(\bm{l}) \rangle = 0 
\end{equation}
From this one can form the quantities:
\begin{equation}\label{vvstar}
\begin{split}
\langle \tilde{v}(\bm{\theta}) \tilde{v}(\bm{\theta'})^{*} \rangle &= \langle v_x v_x' + v_y v_y' \rangle \\ 
&= \int \frac{ d^2 l}{(2 \pi)^2} \left[ C^{EE}_l + C^{BB}_l \right] e^{i\bm{l} \cdot \bm{\Delta\theta}}
\end{split}
\end{equation}
\begin{equation}\label{vv}
\begin{split}
\langle \tilde{v}(\bm{\theta}) \tilde{v}(\bm{\theta}^{'}) \rangle &= \langle v_x v_x' - v_y v_y' + i(v_y v_x' + v_x v_y') \rangle \\ 
&= \int \frac{ d^2 l}{(2 \pi)^2} \left[ C^{EE}_l - C^{BB}_l \right] e^{2i \phi_l} e^{i\bm{l} \cdot \bm{\Delta\theta}}
\end{split}
\end{equation}
In order to form the differential equations for the new $E$ and $B$ power spectra, one simply forms the incremental change (from $\Lambda$ to $\Lambda + \delta \Lambda$) in the first derivative correlator ($\Delta \langle T_i T_j' \rangle$ ) as derived in the previous section.
\begin{equation}\label{changevvstar}
\Delta \langle T_x T_x' + T_y T_y' \rangle = \int \frac{ d^2 l}{(2 \pi)^2} \Delta \left[ C^{EE}_l + C^{BB}_l \right] e^{i\bm{l} \cdot \bm{\Delta\theta}}
\end{equation}
\begin{equation}\label{changevv}
\begin{split}
\Delta \langle T_x T_x' &- T_y T_y' + i(T_y T_x' + T_x T_y') \rangle \\ 
&= \int \frac{ d^2 l}{(2 \pi)^2} \Delta \left[ C^{EE}_l - C^{BB}_l \right] e^{2i \phi_l} e^{i\bm{l} \cdot \bm{\Delta\theta}}
\end{split}
\end{equation}
Now we take the appropriate Fourier transform and take the limit $\delta\Lambda \rightarrow 0$ to get equations for $d C^{EE}_l (\Lambda) /d \ln \Lambda$ and $d C^{BB}_l(\Lambda)/d \ln \Lambda $.  The initial conditions are that $C^{EE}_l = l^2 C^T_l$ and $C^{BB}_l = 0$ at $\Lambda = 0$, which can be deduced from Eq. \eqref{vvstar} and Eq. \eqref{vv} by considering the case $v_i(\bm{\theta}) = T_i(\bm{\beta}_{\Lambda=0}(\bm{\theta})) = T_i(\bm{\theta}) = \partial_i T(\bm{\theta})$.

We now illustrate this derivation for our second order approximation of the system of equations:
\begin{equation}\label{2ndorder1}
\Delta \langle \tilde{T} \tilde{T}' \rangle = -\frac{1}{2} \langle T_i T_j' \rangle \langle \Delta d^i \Delta {d^j} \rangle
\end{equation}
\begin{equation}\label{2ndorder2}
\begin{split}
\Delta \langle T_i T_j' \rangle &= -\frac{1}{2} \langle T_{ia} T_{jb}' \rangle \langle \Delta d^a \Delta {d^b} \rangle \\ 
&\approx -\frac{1}{2} \partial_a \partial_b' \langle T_i T_j' \rangle \langle \Delta d^a \Delta {d^b} \rangle
\end{split}
\end{equation}
Combining Eqns. \eqref{changevv} and \eqref{2ndorder2} one obtains:
\begin{equation}
\begin{split}
&\Delta \int \frac{d^2 l}{(2 \pi)^2} (C^{EE}_l -C^{BB}_l) e^{2i\phi_l} e^{i\bm{l} \cdot \bm{\Delta\theta}} \\
&= -\frac{1}{2} \partial_a \partial_b' \langle v_x v_x' - v_y v_y' + i (v_y v_x' + v_x v_y') \rangle \langle \Delta d^a \Delta d^b \rangle \\
&= -\frac{1}{2} \partial_a \partial_b' \left( \int \frac{d^2 l}{(2\pi)^2} C^{-}_l e^{2i\phi_l} e^{i \bm{l} \cdot \bm{\Delta\theta}} \right) \langle \Delta d^a \Delta d^b \rangle \\
&= -\frac{\delta\Lambda \Lambda C^{\phi}_\Lambda}{2\pi} \int \frac{d \phi_\Lambda}{2\pi} \frac{d^2 l}{(2\pi)^2} (1-e^{i\bm{\Lambda} \cdot \bm{\Delta\theta}}) (\bm{l} \cdot \bm{\Lambda})^2 C^{-}_l e^{2i\phi_l} e^{i\bm{l}\cdot\bm{\Delta\theta}} \\
\end{split}
\end{equation}
where we have defined for convenience $C^{-}_l \equiv C^{EE}_l - C^{BB}_l$.
Taking the Fourier transform $\int d^2 \Delta\theta \, e^{-2i\phi_{l'}} e^{-i \bm{\Delta\theta} \cdot \bm{l'}}$, we get an expression for the change in new spectra:
\begin{equation}
\begin{split}
\Delta \Bigl( C^{EE}_{l'} - &C^{BB}_{l'} \Bigr) = -\frac{\delta\Lambda \Lambda}{2\pi} C^{\phi}_\Lambda \int \frac{d \phi_\Lambda}{2\pi} \Bigl[ C^{-}_{l'} (\bm{l} \cdot \bm{\Lambda})^2 \\
&- C^{-}_{|\bm{l'}-\bm{\Lambda}|} (\bm{\Lambda} \cdot (\bm{l}' -\bm{\Lambda}))^2 e^{2i(\phi_{l'-\Lambda} - \phi_{l'})} \Bigr]
\end{split}
\end{equation}
It is also convenient to define $\tilde{C}^{EE}_l \equiv l^{-2} C^{EE}_l$, $\tilde{C}^{BB}_l \equiv l^{-2} C^{BB}_l$, and $\tilde{C}^{-}_l \equiv \tilde{C}^{EE}_l - \tilde{C}^{BB}_l$ so that $\tilde{C}^{EE}_l$ and $\tilde{C}^{BB}_l$ have the same units as $\tilde{C}^T_l$.
Note that $\text{Re }\left[ e^{2i(\phi_{l'-\Lambda} - \phi_l)} \right] = 2\cos^2(\phi_{l'-\Lambda} - \phi_l) - 1$, and $\cos(\phi_{l'-\Lambda} - \phi_l) = (\bm{l}-\bm{\Lambda}) \cdot \bm{l} /(l |\bm{l}-\bm{\Lambda}|)$.  Then
\begin{equation}
|\bm{l} - \bm{\Lambda}|^2 \text{ Re } \left[ e^{2i(\phi_{l-\Lambda} - \phi_l)}\right] = l^2 F(\Lambda/l, \mu = \cos(\phi_\Lambda-\phi_l))
\end{equation}
with $F(x,\mu) \equiv 1-x^2 -2x\mu + 2 x^2  \mu^2$.  The imaginary part vanishes upon integration.  Finally, taking the limit $\delta\Lambda \rightarrow 0$ the expressions simplify to:
\begin{equation}
\begin{split}
\frac{d}{d\ln\Lambda}\tilde{C}^{-}_l &= \frac{-l^2 \Lambda^4 C^{\phi}_\Lambda}{2 \pi} \int \frac{d \phi_\Lambda}{2\pi} \Bigl[ \tilde{C}^{-}_l \mu^2 \\
&- \tilde{C}^{-}_{|\bm{l}'-\bm{\Lambda}|} [\Lambda/l - \mu]^2 F(\Lambda/l,\mu) \Bigr]
\end{split}
\end{equation}
A similar derivation can be done to find an equation describing the evolution of $\tilde{C}^+_l \equiv \tilde{C}^{EE}_l + \tilde{C}^{BB}_l$.  The result is
\begin{equation}
\begin{split}
\frac{d}{d\ln\Lambda} \tilde{C}^+_l &= \frac{-l^2 \Lambda^4 C^{\phi}_\Lambda}{2 \pi} \int \frac{d \phi_\Lambda}{2\pi} \Bigl[ \mu^2 \tilde{C}^+_l \\
&- \left[ \Lambda/l -\mu\right]^2 G(\Lambda/l,\mu) \tilde{C}^+_{|\bm{l}-\bm{\Lambda}|} \Bigr]
\end{split}
\end{equation}
with $G(x,\mu) \equiv 1 + x^2 -2x\mu$, and $G(\Lambda/l,\mu = \cos(\phi_\Lambda-\phi_l)) = |\bm{l}-\bm{\Lambda}|^2 / l^2$.

Finally, the position-space equation for the correlation function, Eq. \eqref{2ndorder1}, is converted to an equation for $\tilde{C}^T_l (\Lambda)$ by breaking down the first derivative correlator on the right hand side into $E$ and $B$ parts.  The Fourier transform is taken analytically and the limit $\delta\Lambda \rightarrow 0$ is taken.  The final set of coupled equations for the second order solution is:
\begin{equation}\label{final1}
\begin{split}
&\frac{d \tilde{C}^{T}_l(\Lambda)}{d \ln \Lambda}  = -\frac{l^2 \Lambda^4 C^{\phi}_\Lambda}{2\pi} \int \frac{d\phi_\Lambda}{2\pi} \Bigl[ \tilde{C}^{-}_l(\Lambda)\mu^2 \\
&- \tilde{C}^{-}_{|\bm{l}-\bm{\Lambda}|}(\Lambda)  \left[ \frac{\Lambda}{l} - \mu \right]^2 + \tilde{C}^{BB}_l(\Lambda) - G(\frac{\Lambda}{l}, \mu) \tilde{C}^{BB}_{|\bm{l}-\bm{\Lambda}|}(\Lambda) \Bigr] 
\end{split}
\end{equation}
\begin{equation}
\begin{split}
&\frac{d \tilde{C}^{EE}_l(\Lambda)}{d\ln \Lambda}  = -\frac{l^2 \Lambda^4 C^{\phi}_\Lambda}{2\pi} \int \frac{d\phi_\Lambda}{2\pi} \Bigl[ \tilde{C}^{EE}_l(\Lambda) \mu^2 \\ 
&- \left[ \frac{\Lambda}{l} - \mu \right]^2   \Bigl( \tilde{C}^{EE}_{|\bm{l} - \bm{\Lambda}|}(\Lambda) (1- \frac{\Lambda}{l}\mu)^2 + \tilde{C}^{BB}_{|\bm{l}-\bm{\Lambda}|} (\Lambda) \frac{\Lambda^2}{l^2} \nu^2 \Bigr) \Bigr]
\end{split}
\end{equation}
\begin{equation}\label{final3}
\begin{split}
&\frac{d \tilde{C}^{BB}_l(\Lambda)}{d\ln \Lambda} = -\frac{l^2 \Lambda^4 C^{\phi}_\Lambda}{2\pi} \int \frac{d\phi_\Lambda}{2\pi} \Bigl[ \tilde{C}^{BB}_l(\Lambda) \mu^2 \\
&- \left[ \frac{\Lambda}{l} - \mu \right]^2  \Bigl( \tilde{C}^{BB}_{|\bm{l} - \bm{\Lambda}|}(\Lambda) (1- \frac{\Lambda}{l}\mu)^2 + \tilde{C}^{EE}_{|\bm{l}-\bm{\Lambda}|} (\Lambda) \frac{\Lambda^2}{l^2} \nu^2 \Bigr) \Bigr]
\end{split}
\end{equation}
where $\mu = \cos(\phi_\Lambda)$ and $\nu = \sin(\phi_\Lambda)$.  The initial conditions are simply $\tilde{C}^T_l(\Lambda=0) = C^T_l$, $\tilde{C}^{EE}_l(\Lambda = 0) = C^T_l$, and $\tilde{C}^{BB}_l(\Lambda = 0) = 0$.

As compared to the first order equation, Eq. \eqref{approxde}, the role of the lensed temperature power spectrum, $\tilde{C}^T_l(\Lambda)$, in generating the lensing corrections on the right hand side has been replaced by the quantity $\tilde{C}^{EE}_l(\Lambda) -\tilde{C}^{BB}_l(\Lambda)$.  This spectrum generates additional corrections to the lensing and is determined through the second and third equations.  Furthermore, additional corrections are produced through the spectrum $\tilde{C}^{BB}_l$, also evolved from the initial condition through the auxiliary equations.  It is through the additional dynamics of the auxiliary spectra as a function of the cutoff $\Lambda$ that improvements to the first order solution, Eq. \eqref{approxde}, are made.

A similar decomposition can be done for second derivative quantities, in order to generate the equations for a third-order approximation (e.g. Eqs. \eqref{3coupled} - \eqref{3coupled3}).  In this case one can decompose the second derivatives of the temperature on the source plane, $T_{ij}$, into three quantities determined by the spin $\pm 2$ quantities $t(\bm{\theta}) = \frac{1}{2} (T_{xx} - T_{yy}) + i T_{xy}$, $t^*(\bm{\theta}) = \frac{1}{2} (T_{xx} - T_{yy}) - i T_{xy}$, and the scalar $s(\bm{\theta}) =\frac{1}{2} (T_{xx} + T_{yy})$.  In a similar fashion as before, these will be described by four auxiliary power spectra, subject to parity invariance, $C^{E_2 E_2}_l (\Lambda)$, $C^{B_2 B_2}_l (\Lambda)$, $C^{E_2 S_2}_l (\Lambda)$, and $C^{S_2 S_2}_l(\Lambda)$ where the subscript 2 indicates the decomposition of second derivatives to avoid confusion with $E$ and $B$ from the first derivatives (which in turn should not be confused with polarization $E$ and $B$), and $S_2$ denotes the Fourier amplitudes of the scalar $s$.  These derivations can be done mechanically if necessary, in a manner similar to that given above; we will not give the solutions here, but mention this only to illustrate the hierarchical structure of this approach.

\subsection{Interpretation}

It is useful to discuss how this new formulation of the weak lensing calculation differs from previous approaches.  The standard approaches (e.g. \citet{hu00}, \citet{cooray04}) estimate the lensed temperature at a given observed point by expanding the unlensed temperature in a Taylor polynomial in the deflection angle (e.g. Eq. 1), keeping only a few terms.  The accuracy of this expansion will depend on the smoothness of the underlying source temperature field, which depends upon how the fluctuation power is distributed over scales in the intrinsic temperature power spectrum.  The key point is that the unlensed intensity is determined from the lensed image position in a one-step extrapolation depending upon the temperature derivatives evaluated at a single point, i.e. $\bm{\theta} = \bm{\beta}_{\Lambda=0}(\bm{\theta})$.

The main difference in this new method is that the source position is determined in multiple steps, as a function of the cutoff $\Lambda$.  When one integrates over a new shell of the lensing modes with wavenumbers between $\Lambda$ and $\Lambda + \delta\Lambda$, the source position $\bm{\beta}_\Lambda(\bm{\theta})$ changes incrementally through the lens equation.  The first position-space equation for the temperature correlation function tells us to evaluate the source temperature derivatives at the new source position $\bm{\beta}_\Lambda$ in order to determine the incremental modifications to the lensed correlation function.  Hence, by integrating over the lensing spectrum mode by mode, one is continuously updating the current source positions and the current local source gradients, which determine the lensing modifications.

This is analogous to numerical methods for extrapolating a function $f(x)$ to get to $f(x_1)$ from $f(x_0)$.  The simplest thing to do is evaluate
the derivatives at the initial point $x_0$ and extrapolate to get $f(x_1)$
with a Taylor polynomial in one-step.  The accuracy will depend on
the order of the polynomial one uses and how far away $x_0$ is from $x_1$ relative to the scale over which $f(x)$ is varying. More sophisticated methods would use multiple smaller steps in which the derivatives of $f(x)$ are 
evaluated along the trajectory between the points $x_0$ and $x_1$.  Previous methods are analogous to the one-step Taylor expansion calculation.  Our new differential approach is analogous to the latter, adaptive approach, in that the lensing modifications are calculated along the trajectory on the source plane between the observed position $\bm{\beta}_{\Lambda=0}(\bm{\theta}) = \bm{\theta}$ and the final source temperature position $\bm{\beta}_{\Lambda = \infty}(\bm{\theta})$.

The first equation, Eq. \eqref{topeq}, which determines the lensed temperature correlation function as a function of the cutoff, is exact in the limit of the flat-sky and in the limit that we take the lensing shell increment $\delta\Lambda \rightarrow 0$.  This is a consequence of the fact that we did not expand in the full deflection angle, but only in the differential increment of the deflection field containing the new modes.  However, in order to evaluate the right hand side of the differential equation, one must evaluate the first derivative correlator at the new source positions.  The implication is that we have to evaluate the second derivatives at the changing source position $\bm{\beta}(\Lambda)$.  Then to update the changing second derivative correlator we have to calculate third derivative correlators, and so on.  This leads to the hierarchy of equations for the temperature correlator and temperature derivative correlators.  We can terminate and close the hierarchy at any level by approximating a derivative on the source plane with a derivative on the observed image plane, at the cost of neglecting terms representing the relative distortion of these derivatives through the dimensionless convergence and shear, which are presumed to be small $(|\kappa|, |\gamma_1|, |\gamma_2| < 1)$ in the limit of weak lensing.  

This brings us to a second major distinction of this method.  Whereas the previous methods are essentially perturbative in the full deflection angle, so that the truncation errors are simply those attributed to high-order terms in the deflection angle, in this new formulation the approximation is based upon the smallness of elements of the distortion tensor, which are derivatives of the deflection angle, in weak lensing.  Intuitively, the key point we emphasize here is that the RMS magnitude of the deflection at a single point is not the physically relevant quantity in determining the lensing modification to the two-point function or the power spectrum.  What is physically important  is the difference between the deflection angles at two points, which is determined by the dimensionless second derivatives of the projected potential through the convergence and shear.  Hence, in the context of the weak lensing calculation, it is more appropriate to approximate about the dimensionless elements of the distortion tensor rather than the deflection angle.  Furthermore, this approximation depends primarily on the weakness of the distortion tensor elements rather than the properties of the temperature fluctuations being lensed.

\section{Numerical Implementation and Results: 21 cm Fluctuations}
\label{clresults}

The numerical implementation of these systems of equations is simple.  In order to implement the second order approximation, for example, to compute the lensing effect on a power spectrum sampled at $N$ points, one simply creates a vector of $3N$ elements.   This is initialized to the values of the unlensed temperature spectrum, $C^T_l$ for the first $N$ values, the initial values of $\tilde{C}^{EE}_l$ (which are $C^T_l$) for the next $N$, and then the initial values of $\tilde{C}^{BB}_l$, which are zero for the final $N$ slots.  This vector is then fed into an adaptive Runge-Kutta 4th and 5th order algorithm, and an adaptive Gauss-Kronrod scheme is used to evaluate the one-dimensional integrals on the right hand sides of the equations with the help of smooth interpolations between the sample points.  Since the intrinsic power spectra are very smooth, these integrations are simple.

We apply our first and second order sets of lensing differential equations to the case of 21 cm fluctuations.  The fluctuations in the 21 cm emission at redshift $z$ as seen against the CMB are described by:
\begin{equation}
\begin{split}
\delta T(\nu) = &23 \text{ mK} \left[\left(\frac{0.15}{\Omega_M h^2}\right) \left(\frac{1+z}{10}\right) \right]^{1/2} \\
&\times \left(\frac{T_S - T_{CMB}}{T_S}\right) \left(\frac{\Omega_b h^2}{0.02}\right) x_H (1 + \delta)
\end{split}
\end{equation}
(e.g. \citet{zfh04}).  In the simplification that the spin temperature $T_S$ is coupled to the kinetic temperature $T_K$ of the gas that is heated far above the CMB temperature, $T_S \gg T_{CMB}$, the fluctuations are independent of the spin temperature.  Then the intensity fluctuations are due to the fluctuations in the quantity $\psi = x_h (1+\delta)$, where $x_H$ is the neutral hydrogen fraction, and $\delta$ the matter density contrast.  We have ignored additional sources of fluctuations, e.g. peculiar velocities, for simplicity.

\begin{figure}[t]
\centering
\includegraphics[angle=90,scale=0.4]{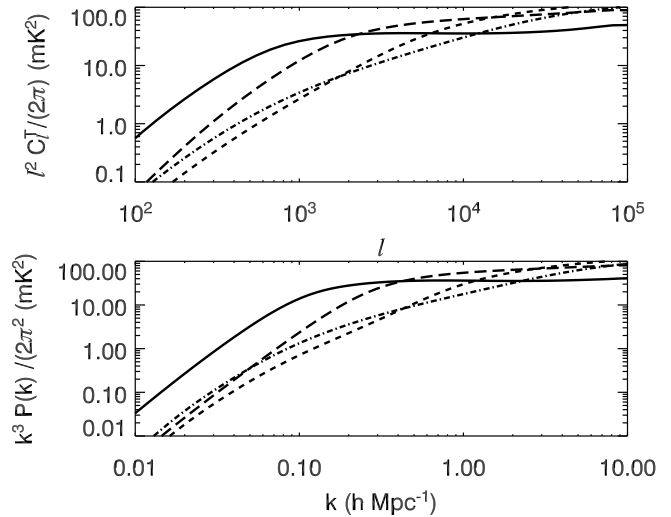}
\caption{\label{displayspectra} Model power spectra of 21 cm brightness temperature fluctuations used in this paper.  Angular power spectra versus angular multipole are shown in the top panel.  In the bottom panel, the three-dimensional power spectrum is plotted versus comoving wavenumber.  The curves correspond to $z = 12$ (solid), $z = 13$ (long dash), $z = 15$ (short dash), and $z = 18$ (dash dot).}
\end{figure}

We have taken semi-analytic models of the expected 21 cm fluctuation power spectra at several redshifts.  These correspond to a sample model of a single reionization history as described in \citet{fzh04a}.  This model is described by an ionization efficiency factor $\zeta = 40$, and we use the resulting temperature power spectra at redshifts of $z = 12$, $13$, $15$, and $18$, corresponding to mean neutral fractions $\bar{x}_H = 0.25$, $0.5$, $0.8$, and $0.96$.  These power spectra are shown in Fig. \ref{displayspectra}.  For a lower ionization efficiency, reionization will occur at lower redshifts; however, the power spectra remain similar for a given neutral fraction \citep{fzh04a}.  Hence, one expects that the lensing of 21 cm spectra for different models at different redshifts will be qualitatively similar, modulated by the changes in the lensing power spectra to those redshifts (Fig. \ref{convergenceplot}).

\begin{figure}[t]
\centering
\includegraphics[angle=90,scale=0.4]{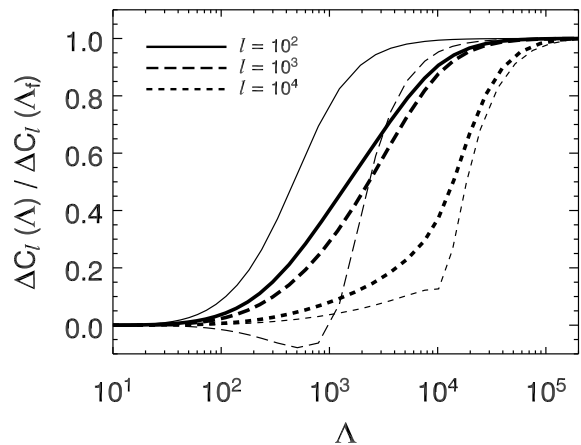}
\caption{\label{diffeqsoln}Solutions to the system of differential equations Eqns. \eqref{final1} - \eqref{final3} as a function of the lensing cutoff $\Lambda$ at representative scales.  The curves describe $\Delta C_l(\Lambda)/C_l \equiv (\tilde{C}^T_l(\Lambda) - C^T_l)/C^T_l$ scaled to the final value at $\Lambda = \Lambda_{\text{final}} = 3 \times 10^5$, at multipole values $l = 10^2$ (solid), $l = 10^3$ (long dash), and $l = 10^4$ (short dash) for the z = 12 (thin) and z = 15 (thick) model power spectra.The small scale modes at $l \sim 10^4$ where the power spectra are relatively near scale-invariant are comparatively unaffected by lensing modes $\Lambda <l$, and mostly lensed by deflection modes $\Lambda \gtrsim l$.  At intermediate and larger scales, temperature modes are affected by a window of lensing modes wavenumbers $\Lambda \sim 10^2 \text{ to } 10^4$.   }
\end{figure}

\begin{figure}[t]
\centering
\includegraphics[angle=90,scale=0.4]{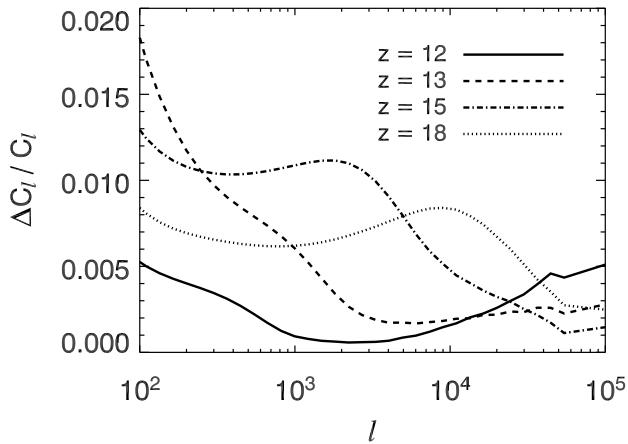}
\caption{\label{angularlensing}Gravitational lensing effects on 21 cm angular power spectra.  The fractional modifications of the angular power spectra, $\Delta C_l /C_l \equiv (\tilde{C}^T_l - C^T_l )/C^T_l$ are plotted as a function of angular scale $l$. The curves correspond to the z = 12 (solid), z = 13 (dashed), z = 15 (dashed-dot) and z = 18 (dotted) models as described in the text.  The lensing modifications towards the end of reionization are much smaller on small angular scales due to the flattening of the power spectra by the distribution of bubbles.  At earlier epochs, the small-scale lensing is larger due to the steeper slope.  The gravitational lensing effect on large scales is larger due to the steeper slopes of the power spectra.}
\end{figure}

We apply our first and second order lensing systems of equations to calculate the effect on the angular power spectrum.  Solutions are shown in Figure \ref{diffeqsoln}.  The fractional change in the temperature power spectrum is shown in Figure \ref{angularlensing}.  Comparing the solutions to the first and second order approximations, we have found that the difference is typically of order $1\%$ of the lensing effect.  This demonstrates that our calculations are convergent and are a complete description of the lensing effect from degree scales to arcsecond scales.  There is no evidence that the calculations break down at any scale, despite the dominance of small scale power.

The results show that the lensing modifications are small on scales $l \gtrsim 1000$ for late-time spectra ($z = 12, 13$).
The intrinsic power spectra approach a scale-invariant slope at these small angular scales.  We have seen that the effect of lensing on a near scale-invariant spectrum is small.  For lensing modes that are slowly varying compared to a temperature mode $l$ there is no effect.  Since a mode in a scale-invariant spectrum is not affected by longer-wavelength lensing modes,  the temperature modes at high-$l$ are unaffected by most of the lensing power at $\Lambda < l$; the lensing effect is determined by the convergence power near the scale $l$.  This can be seen in Figure \ref{diffeqsoln}.

Note that the largest relative lensing effect occurs on the rising slope of the temperature power spectrum $l^2 C^T_l /(2\pi)$, with the most pronounced effect occurring on the steepest slope.  The implication is that, as reionization occurs, and the bubble feature ``shoulder'' is imprinted on the angular power spectrum on scales $l \sim 10^3$ to $10^4$, the relative lensing effect on the scale of the feature and on smaller scales diminishes as the spectrum flattens out to near scale-invariant form.  

\section{Lensing effects on the Three-dimensional Power Spectrum}
\label{3dlensingsection}

Just as weak lensing will modify the angular power spectra of 21 cm fluctuations, it will also affect the observed three-dimensional power spectrum of the intensity, $\tilde{P}_I(\bm{k})$.  The isotropic nature of the random field of fluctuations at a particular epoch implies that the intrinsic power spectrum will be approximately spherically symmetric, $P_I(\bm{k}) = P_I(k)$; however asymmetries are introduced by velocity field distortions (Barkana \& Loeb 2005).  The three-dimensional Fourier tranform of the power spectrum is the three-dimensional autocorrelation function of the intensities, $\langle I(\bm{r}) I(\bm{r}') \rangle$.  It is clear that the observed positions of the intensities will be displaced from the true source positions through the effects of weak gravitational lensing. In this section, we will show that gravitational lensing will modify the observed power spectrum, and also will introduce angular anisotropies into $\tilde{P}_I(\bm{k})$.

We consider the scenario in which an image cube of intensities in a three-dimensional region is constructed with observations all at approximately the same epoch.  In principle, another anisotropy will be introduced due to the evolution of the intensity power spectrum with time.  Thus, the image cubes must be constructed over a redshift range that is narrow compared to the rapidity with which reionization occurs at a given epoch.  

\begin{figure}[t]
\includegraphics[angle=90,scale=0.4]{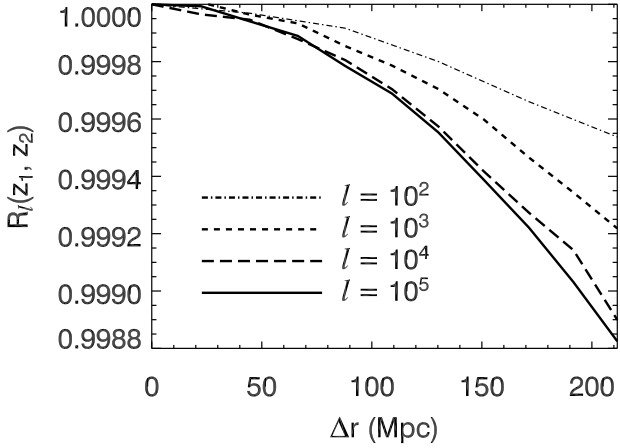}
\centering
\caption{\label{crossconvplot} Cross-correlation coefficients of the lens convergence, Eq. \ref{crosscoeff}.  The curves represent the coefficient $R_l (z_1, z_2)$ between $z_1 = 10$ and redshifts $z_2$ between $z_2 = z_1$ and $z_2 = 11$.  This range corresponds to a radial width of $\Delta r \equiv r(z_2) - r(z_1) \approx 210 \text{ Mpc}$.  The different curves represent a range of multipoles from $l = 10^2$ to $l = 10^5$.  The lens convergences decorrelate only by $\lesssim 0.1\%$ over radial widths of 200 Mpc.}
\end{figure}

In general the lensing power to each redshift (frequency) plane of the image cube will be different.  The variation of the lensing effect between two redshift planes can be characterized by the convergence cross-power spectrum between the two redshifts, $z_1 < z_2$:
\begin{equation}
C^{\kappa_1\kappa_2}_l = \frac{9 H_0^4 \Omega_M^2}{4 c^4} \int_0^{r(z_1)} dr \frac{W(r, r_1) W(r, r_2)}{a^2(r)} P_{\delta}\left(\frac{l}{\chi(r)}, r \right)
\end{equation}
The cross-correlation coefficient is then defined as:
\begin{equation}\label{crosscoeff}
R_l(z_1,z_2) = \frac{C^{\kappa_1 \kappa_2}_l}{\sqrt{C^{\kappa_1}_l C^{\kappa_2}_l}}
\end{equation}
where $\kappa_1 \equiv \kappa(z_1)$ and $\kappa_2 \equiv \kappa(z_2)$.
We have calculated the cross-correlation for typical redshifts in this regime over the expected radial widths of the image cubes, on the order of $\sim 100$ Mpc. In Figure \ref{crossconvplot}, the cross-correlation coefficient of the convergence between $z = 10$ and neighboring redshift planes up to $z = 11$ is shown for several multipoles between $l = 10^2 \text{ and } 10^5$.  The coefficient is negligibly different from unity over comoving radial differences $\Delta r \sim 200 \text{ Mpc}$.  The decorrelation of the lens convergences is only $\lesssim 0.1\%$ over this range.  This is primarily because the lensing efficiency is greatest for lens planes at substantially lower redshifts of order $z \sim 1$, so that small differences in the source distance make negligible differences to the total lensing effect.

Consider an image cube observed centered at redshift $z_*$, at a central radial distance $r_*$ and comoving angular diameter distance $D_*$.  Assume that the radial width of the box is small enough so that for each pair of constituent planes at different frequencies (redshifts), the cross-correlation of the lensing convergences is near unity.  We have shown that this is an excellent approximation for radial widths up through hundreds of Mpc.  The deflection field traversed by light from a diffuse source generally depends on its distance and redshift in the three-dimensional volume: $\bm{\delta\theta}(\bm{\theta}; z)$.  This approximation allows one to make the simplification that the deflection field seen by each of the multiple frequency-planes comprising the box is the same: $\bm{\delta\theta}(\bm{\theta}, z_1) \approx \bm{\delta\theta}(\bm{\theta}, z_2) \approx \bm{\delta\theta}(\bm{\theta}, z_*) \equiv \bm{\delta\theta}(\bm{\theta})$. The statistics of the deflection fields can thus be described by a single power spectrum $C^{\phi\phi}_l(z_*)$.

The apparent comoving transverse displacement due to lensing of an observed ray $\tilde{I}(\bm{r}) = \tilde{I}(\bm{\theta} D(z), r_z)$ from the plane at redshift $z_1$ is $\Delta\bm{r_\perp}(\bm{\theta} D(z_1), z_1) = D(z_1) \bm{\delta\theta}(\bm{\theta},r_{z_1}) )$.  We will make the further simplification that the proportionality factor, the comoving angular diameter distance, is approximately the same for each plane in the box, and equal to the central distance.  Thus, $\Delta\bm{r_\perp}(\bm{\theta}D(z_1), z_1) \approx D(z_*) \bm{\delta\theta}(\bm{\theta}, r(z_1)) \approx D_* \bm{\delta\theta}(\bm{\theta},z_*)$. This approximation will be good so long as the box is not too long in the radial dimension.  For example, in a box of length $\sim$ 100 Mpc at a typical distance $D_* \sim 10^4$ Mpc, the variation of $D(z)$ will only be of order a few percent across the box in the radial direction.

The intrinsic intensity field due to fluctuations within the box is described by
\begin{equation}
\begin{split}
I(\bm{r}) &= \int \frac{d^3 k}{(2\pi)^3} I(\bm{k}) e^{i\bm{k}\cdot{\bm{r}}} \\
&= I(\bm{\theta} D(z), r_z) = \int \frac{d^2 k_\perp}{(2\pi)^2} \frac{d k_\parallel}{2\pi} I(\bm{k}) e^{i \bm{k_\perp} \cdot \bm{\theta} D(z)} e^{i k_\parallel r_z} 
\end{split}
\end{equation}
where we have split up the comoving coordinate $\bm{r}$ into components perpendicular and parallel to the line of sight.  Here, $D(z)$ denotes the co-moving angular diameter distance to redshift $z$,  $r_z$ is the comoving distance to redshift $z$, and $\bm{k}$ denotes comoving wavevectors, also split up into transverse and parallel components.  The ensemble average over intensity fluctuations of the correlation function is 
\begin{equation}\label{3dcorrelation}
\langle I(\bm{r}) I(\bm{r}') \rangle_I = \int \frac{d^3 k}{(2\pi)^3} P_I (\bm{k}) e^{i\bm{k}\cdot (\bm{r}-\bm{r}')}
\end{equation}
where $P_I(k)$ denotes the intrinsic three-dimensional power spectrum.  Under the simplifications stated above, gravitational lensing will remap the observed intensity $\tilde{I}(\bm{r}) = \tilde{I}(\bm{\theta} D,r_z) = I( D\bm{\theta}+D_*\bm{\delta\theta}(\bm{\theta}), r_z)$.  The observed, lensed correlation function is modified as:
\begin{equation}\label{pobvious}
\begin{split}
&\langle \tilde{I}(\bm{r}) \tilde{I}(\bm{r}')\rangle_{I,\phi} = \langle I(D(\bm{\theta} + \bm{\delta\theta}), r_z) I(D' (\bm{\theta'} + \bm{\delta\theta'}),r_z') \rangle_{I,\phi}\\
&= \langle I(D \bm{\theta}, r_z) I(D' \bm{\theta}' + D'\bm{\delta\theta}' - D\bm{\delta\theta}, r_z' \rangle_{I, \phi} \\
&\approx \langle I(\bm{r}) I(\bm{r}') \rangle_I + \langle I(\bm{r}) \frac{1}{2} {\nabla_\perp^a}' {\nabla_\perp^b}' I(\bm{r'}) \rangle_I D_*^2 \langle \Delta \delta\theta^a \Delta \delta\theta^b \rangle_\phi
\end{split}
\end{equation}
where we have expanded to second order in the relative deflection angle.  Here $\bm{\nabla_\perp}$ denotes the gradient along the transverse direction (on the sky), and as before $\Delta\bm{\delta\theta} \equiv \bm{\delta\theta'} - \bm{\delta\theta}$.  With $\nabla_\perp^a \nabla_\perp^b I(\bm{r}) = \int d^3 k / (2\pi)^3 \, -k_\perp^a k_\perp^b I(\bm{k}) \exp(i\bm{k}\cdot\bm{r})$, and defining the lensed, observed power spectrum, $\tilde{P}_I(\bm{k})$ with $\langle \tilde{I}(\bm{r}) \tilde{I}(\bm{r}')\rangle_{I,\phi} = \int d^3 k / (2\pi)^3 \, \tilde{P}_I(\bm{k}) \exp(i\bm{k}\cdot{\bm{\Delta r}})$, we find upon taking the three-dimensional transform,
\begin{equation}\label{pkhulensed}
\begin{split}
\tilde{P}_I(\bm{k}_{\perp}, k_{\parallel}) &- P_I(k) = -\int \frac{d^2 l}{(2\pi)^2}  C^{\phi}_l \Bigl\{ P_I(\bm{k}_{\perp}, k_{\parallel}) (D_* \bm{k} \cdot \bm{l})^2 \\ 
&- P_I(\bm{k}_{\perp} - \bm{l}/D_*, k_{\parallel}) \left[ \bm{l} \cdot (D_* \bm{k}_{\perp} - \bm{l}) \right]^2 \Bigr\}
\end{split}
\end{equation}
where in the case of spherical symmetry of the intrinsic power spectrum, $P(k_\perp, k_\parallel) = P([k_\perp^2 + k_\parallel^2]^{1/2})$.  This equation is structurally similar to that of the first order lensing effect on an angular power spectrum, except here we have a three-dimensional intrinsic power spectrum.  It is easy to see from this equation how anisotropy is introduced into the power spectrum through lensing; the right-hand side of the equation depends separately on the transverse component, $k_{\perp}$.  It is also apparent that under the approximations,  the lens effect on a particular mode with line-of-sight wavenumber $k_{\parallel}$  only involves mixing with other modes with the same line-of-sight component.  Thus, we may consider the case of three-dimensional lensing as analogous to the case of two-dimensional lensing on effective power spectra $P(k_\perp, k_\parallel)$ as a function of $k_\perp$ for each fixed $k_\parallel$. 

The structural similarity of this lensing formula to that of the two-dimensional case (Eq. \eqref{husoln}) suggests that the differential techniques developed in \S \ref{generalapproach} can be readily adapted to this case in order to calculate higher order corrections and to verify convergence.  This derivation is done in a nearly identical fashion.  We can simply make the replacement $\tilde{C}_l (\Lambda) \rightarrow P_\Lambda(k_\perp, k_\parallel)$ to denote the power spectrum lensed by lensing modes with wavenumber less than $\Lambda$, and substitute $l \rightarrow k_\perp D$.  The resulting differential equation at first approximation are:
\begin{equation}\label{pdiffeq1}
\begin{split}
&\frac{d \tilde{P}_\Lambda(\bm{k})}{d \ln \Lambda}  = \frac{-\Lambda^2 C^{\phi}_\Lambda}{2\pi} \int \frac{d\phi_\Lambda}{2\pi} \Bigl\{ \tilde{P}_\Lambda(k_\perp, k_\parallel) (D_* \bm{k_\perp} \cdot \bm{\Lambda})^2 \\
&- \tilde{P}_\Lambda(|\bm{k}_\perp - \bm{\Lambda}/D_*| ,k_\parallel)\left[ \bm{\Lambda} \cdot (D \bm{k}_\perp -\bm{\Lambda}) \right]^2  \Bigr\}
\end{split}
\end{equation}
with $P_{\Lambda = 0}(k_\perp, k_\parallel) = P(k_\perp, k_\parallel)$.
The analogous system at the second approximation is:
\begin{equation}\label{pdiffeq2a}
\begin{split}
\frac{d\,\tilde{P}_\Lambda(\bm{k})} {d \ln \Lambda} &=  \frac{-\Lambda^2 C^{\phi}_\Lambda}{2\pi} \int \frac{d \phi_\Lambda}{2 \pi} \biggl( (D_*\bm{k}_\perp \cdot \bm{\Lambda})^2 \tilde{P}^{-}_\Lambda (k_\perp, k_\parallel) \\
&- \left[\bm{\Lambda} \cdot (\bm{\Lambda} - D_*\bm{k_\perp}) \right]^2 \tilde{P}_\Lambda^{-}(|\bm{k_\perp}  - \frac{\bm{\Lambda}}{D_*}|, k_\parallel) \\
&+  \Lambda^2 k_\perp^2 D_*^2 \tilde{P}^{BB}_\Lambda (k_\perp, k_\parallel) \\
&- \Lambda^2 |\bm{\Lambda} - D_* \bm{k}_\perp|^2 \tilde{P}_\Lambda^{BB}(|\bm{k}_\perp - \bm{\Lambda}/D_*|, k_\parallel) \biggr) \\
\end{split}
\end{equation}
\begin{equation}\label{pdiffeq2b}
\begin{split}
&\frac{d\,\tilde{P}^{BB}_\Lambda(\bm{k})} {d \ln \Lambda}  = \frac{-\Lambda^2 C^{\phi}_\Lambda}{2\pi} \int \frac{d \phi_\Lambda}{2\pi} \Bigl\{ (D_* \bm{k}_\perp \cdot \bm{\Lambda})^2 \tilde{P}^{BB}_\Lambda(k_\perp, k_\parallel) \\
&- \bigl[\bm{\Lambda} \cdot (D_* \bm{k}_\perp - \bm{\Lambda})  \bigr]^2 \Bigl( \tilde{P}^{BB}_\Lambda(|\bm{k} - \frac{\bm{\Lambda}}{D_*}|, k_\parallel) (1 - \frac{\Lambda}{k_\perp D_*} \mu)^2 \\
&+ \tilde{P}^{EE}_\Lambda(|\bm{k_\perp} - \frac{\bm{\Lambda}}{D_*}|, k_\parallel) \frac{\Lambda^2}{k_\perp^2 D_*^2} \nu^2 \Bigr)\Bigr\}
\end{split}
\end{equation}
\begin{equation}\label{pdiffeq2c}
\begin{split}
&\frac{d\,\tilde{P}^{EE}_\Lambda(\bm{k})} {d \ln \Lambda}  = - \frac{\Lambda^2 C^{\phi}_\Lambda}{2\pi} \int \frac{d \phi_\Lambda}{2\pi} \Bigl\{ (D_* \bm{k}_\perp \cdot \bm{\Lambda})^2 \tilde{P}^{EE}_\Lambda(k_\perp, k_\parallel) \\
&- \bigl[\bm{\Lambda} \cdot (D_* \bm{k}_\perp - \bm{\Lambda})  \bigr]^2 \Bigl( \tilde{P}^{EE}_\Lambda(|\bm{k} - \frac{\bm{\Lambda}}{D_*}|, k_\parallel) (1 - \frac{\Lambda}{k_\perp D_*} \mu)^2 \\
&+ \tilde{P}^{BB}_\Lambda(|\bm{k_\perp} - \frac{\bm{\Lambda}}{D_*}|, k_\parallel) \frac{\Lambda^2}{k_\perp^2 D_*^2} \nu^2 \Bigr)\Bigr\}
\end{split}
\end{equation}
where $\mu = \cos(\phi_\Lambda - \phi_{k_\perp})$ and $\nu =  \sin(\phi_\Lambda - \phi_{k_\perp})$, with $\tilde{P}_{\Lambda=0}(k_\perp, k_\parallel) = \tilde{P}^{EE}_{\Lambda=0}(k_\perp, k_\parallel) = P(k_\perp, k_\parallel)$, $\tilde{P}^{BB}_{\Lambda=0} = 0$ as initial conditions.  In the limit that the first approximation, Eq. \eqref{pdiffeq1}, is sufficient, and that lensing modifications are small, then the solution, Eq. \eqref{pobvious}, is a sufficient description of the lensing effect, as one may replace $\tilde{P}_\Lambda(\bm{k}) \approx P(\bm{k})$ on the right hand side.

\begin{figure}[t]
\centering
\includegraphics[angle=90,scale=0.4]{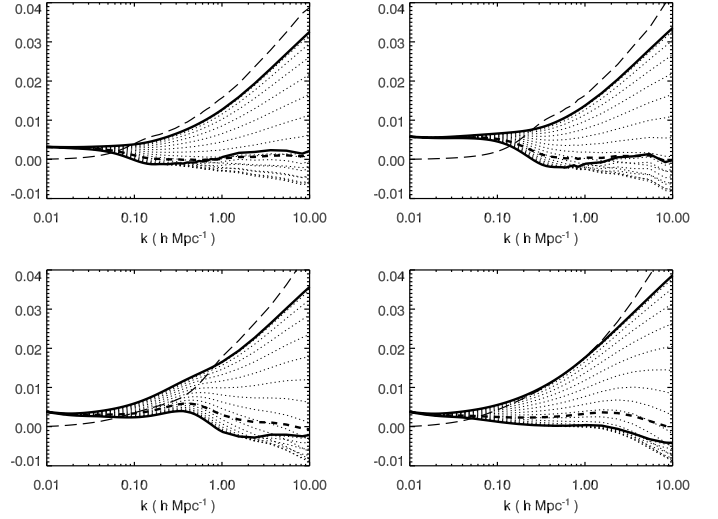}
\caption{\label{3dlensing}Lensing effect on the three-dimensional power spectrum $P(\bf{k})$.  The curves represent the fractional lensing effect, $\Delta^P(k,\alpha) \equiv (\tilde{P}(\bm{k}) - P(\bm{k}))/P(\bm{k})$, on the power spectrum of modes with a given magnitude $k$ and orientation $\alpha$, as described in the text. These pertain to the models at z = 12 (top left), z = 13 (top right), z = 15 (bottom left), z = 18 (bottom right) as described in the text.  The upper solid line is for purely line-of-sight wavevectors ($\alpha = 0$), the lower solid line represents purely transverse wavevectors ($\alpha = \pi/2$), and the thin dotted lines represent intermediate orientations at a given $k$, $\alpha \in (0,\pi/2) $.  The short dashed line is the spherical average of the fractional lensing (over all orientations $\alpha$). The long dashed line is a measure of the anisotropy due to lensing equal to the maximum minus the minimum of $\Delta^P(k,\alpha)$ for a given $k$. }
\end{figure}

In Figure \ref{3dlensing} we show the solutions to these equations for the models of intrinsic 21 cm power spectra as discussed \S \ref{clresults}.  Here we have assumed for simplicity that the intrinsic power spectrum is spherically symmetric and ignored any components, such as velocity field distortions, that may vary with angle $\cos\alpha = k_\parallel/k$.  Plotted is the lensed fraction of the power spectrum, which is defined as
\begin{equation}
\Delta^P(\bm{k}) = \frac{\tilde{P}_I(\bm{k}) - P_I(\bm{k})}{P_I(\bm{k})},
\end{equation}
as a function of the wavenumber $k$ for several orientations (dotted lines) of the wavevector $\bm{k} = (k_\parallel,k_\perp) \equiv (k \cos\alpha, k \sin\alpha)$, where $\alpha$ is the angle between the line of sight and $\bm{k}$.  These orientations range from fully transverse on the sky ($\alpha = \pi/2$) to aligned with the line-of-sight ($\alpha = 0$).  In principle, this treatment is not appropriate for nearly line-of-sight modes, i.e. those with very small or no transverse component on the sky, $k_\perp \sim 0$, because we have adopted a flat-sky formalism, and sky curvature effects will be important for these modes.  In practice, these modes will not be observeable due to small finite survey areas.  
This limit may be described, at each  $k = (k_\perp^2 + k_\parallel^2)^{1/2}$, by the lensed fraction for modes with $l \sim k_\perp D = 100$.  This limit is only appreciably different from the lensed fraction of line-of-sight modes for modes with $k \sim 0.01 \text{ } h \text{ Mpc}^{-1}$; on small scales $k \gg 0.01 \text{ } h \text{ Mpc}^{-1}$ this limit is very close to the line-of-sight mode.  
We also compute the spherically averaged lensed fraction, $\int_{-1}^{1} d\cos\alpha\, \Delta^P(k,\alpha) / 2$, (thin dashed line), and the anisotropy ratio, which, defined as
\begin{equation}
R^P(k) = \frac{1 + \max\limits_{0 < \alpha < \pi/2} \Delta^P(k,\alpha)}{1 + \min\limits_{0 < \alpha < \pi/2} \Delta^P(k,\alpha)}
\end{equation}
quantifies the amplitude of the variation in lensing over all orientations, and measures the width of the filled regions in Fig. \ref{3dlensing}.  As in the two-dimensional case, we find that the difference between the second and first order solutions is very small, and hence higher order terms are negligible.

The most interesting aspect of these results is the variation of the gravitational lensing effect as one rotates the wavevector $\bm{k}$ about a fixed axis in the plane of the sky.  On large scales, $k \lesssim 0.1 \text{ } h \text{ Mpc}^{-1}$, the lensing fraction is roughly isotropic and of order fractions of a percent.  On smaller scales, $k \gtrsim 0.1 \text{ } h \text{ Mpc}^{-1}$, the lensed fraction varies significantly with polar angle $\alpha$.  The high inclination modes with small $\alpha$  have the largest lensed fraction, whereas transverse modes have near zero or slightly negative lensed fraction. It is not necessarily true on small scales that the smallest lensing effect occurs for purely transverse modes; this will generally depend on the small scale shape of the power spectrum.  Meanwhile, the spherically averaged lensed fraction is consistently small, of order several tenths of a percent, over several decades in scale in $k$.  The anisotropy ratio on small scales can approach $R^P \sim 1\% $ at $k \gtrsim 1 \text{ } h \text{ Mpc}^{-1}$.

One might expect intuitively that, since gravitational lensing introduces deflections transverse to the line of sight, only the power in transverse modes will be affected significantly.  In fact, as is evident in Figure \ref{3dlensing}, the opposite is true.
The described behavior of the lensing effect can be understood through the shape of the power spectrum.  In the familiar case of the lensing of the two-dimensional power spectrum, the quantity of interest is the power spectrum of the temperature gradient on the sky, $l^2 C^T_l$.  Analogously, in the three-dimensional case, what is important is the power spectrum of the temperature gradient in the plane of the sky, that is $k_\perp^2 P(k_\perp, k_\parallel)$.  In the three-dimensional case, one can consider the total lensing effect as the set of lensing modifications of effective power spectra in the transverse wavenumber $k_\perp$, each labelled by $k_\parallel$.  For a fixed $k_\parallel$, these effective (intrinsic) power spectra behave as 
\begin{equation}\label{regimecases}
k_\perp^2 P(\bm{k}) =  \begin{cases} k_\perp^2 P(k_\perp) & \text{if $k_\perp \gg k_\parallel$}, \\ k_\perp^2 P(k_\parallel) \sim k_\perp^2 & \text{if $k_\perp \ll k_\parallel$}. \end{cases}
\end{equation}
We see that in the case of a nearly transverse mode ($k_\perp \gg k_\parallel$), the effective power spectrum is locally simply that of the underlying power spectrum $k^2 P(k)$.  This is to be contrasted with the case of the highly inclined mode ($k_\perp \ll k_\parallel$) where locally the power spectrum is characteristic of Poisson fluctuations: $P(k_\perp, k_\parallel)$ is approximately constant versus $k_\perp$.  The effect of this is to steepen the slope on scales $k_\perp \ll k_\parallel$, thereby enhancing the lensed fraction.  We have seen in the case of the two-dimensional power spectrum that the Poisson approximation $C^T_l \sim \text{ constant}$ leads to lensing by a positive constant determined by the total convergence power.  This result applies here approximately for highly inclined modes.  Specifically, for $k_\perp^2 P(\bm{k}) \sim k_\perp^2$, we have, from Eq. \eqref{pkhulensed},
\begin{equation}
\Delta^P(k_\perp, k_\parallel) \approx 4 \int \frac{d^2 l}{(2\pi)^2} C^{\kappa\kappa}_l = 4 \langle \kappa^2 \rangle
\end{equation}
so that the lensed fraction is a measure of the total power in convergence.  For realistic spectra, this approximation is only valid at transverse scale $k_\perp^0$ for lensing modes up to a cutoff $\Lambda$ such that $\Lambda + k_\perp^0 \ll k_\parallel$, since the effective transverse spectrum will be shallower than Poisson at small scales, i.e. the large $k_\perp$ limit of Eq. \ref{regimecases}.  Thus, the above simplification serves approximately as an upper bound.

\begin{figure}[t]
\centering
\includegraphics[angle=90,scale=0.4]{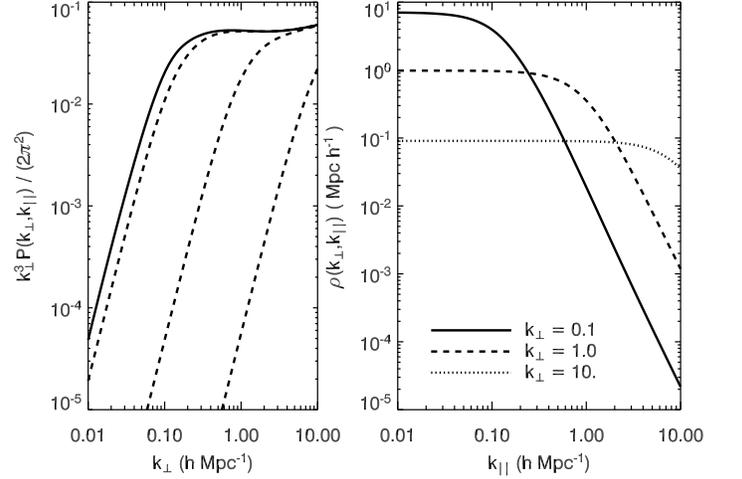}
\caption{\label{rho} Left:  Effective transverse power spectra in $k_\perp$ for the $z = 12$ model.  The solid line is simply the three-dimensional spectrum $k^3 P(k) /(2\pi^2)$, or equivalently the effective transverse spectrum for $k_\parallel = 0$, e.g. fully transverse modes.  The dashed lines are the effective spectra for (left to right) $k_\parallel = \text{0.1, 1.0, and 10 h Mpc}^{-1}$.  The normalization is arbitrary.  For highly inclined modes, i.e. $k_\perp \ll k_\parallel$, the effective spectrum that is subject to lensing is steepened compared to the three-dimensional spectrum (solid line) into a Poisson form.  Right: Weighting function, $\rho(k_\perp, k_\parallel)$ as a function of $k_\parallel$.  Because of the decline in the power spectrum $P(k)$ on small scales, the lensing of the inclined ($\alpha \gtrsim \pi/4$ or $k_\parallel > k_\perp$) three-dimensional modes are discounted relative to transverse modes in generating the lensing effect on the angular power spectrum.}
\end{figure}

For a given $k$, i.e. a vertical slice of Figure \ref{3dlensing}, the lensing of the nearly transverse modes can be considered as the lensing of the intrinsic spectrum $P(k)$ at a high $k_\perp \approx k$, while the lensing of the highly inclined modes can be considered approximately as the lensing of a Poisson spectrum at low $k_\perp$.  Hence, the intuition that only transverse modes are lensed is incorrect: physically one must consider the effective transverse power spectrum presented to the gravitational lens.  To make this clear, Figure \ref{rho} shows the effective power spectra in $k_\perp$ for several values of $k_\parallel$.  One can also see why the anisotropy due to lensing is small on large scales, and grows with $k$.  For small $k$, the effective power spectra are not appreciably different as one changes the component $k_\parallel$, furthermore the lensing effect on the rising slope is dominated by contributions from smaller scale temperature modes, and thus relatively independent of rotating the particular components $(k_\parallel, k_\perp)$.  On small scales or large $k$, the $k \approx k_\parallel$ effective spectrum is significantly modified compared to the $k = k_\perp$ spectrum, and the angular range, $0 \gtrsim \alpha \gtrsim \pi/2$, spans a wider difference in $k_\perp$.

\subsection{Relation between 3D and 2D Lensing}

One may observe that the lensed fraction of the three-dimensional power spectrum can become relatively large ($\gtrsim 1\%$) for some modes at small scales, while the two-dimensional lensing at equivalent angular scales is comparatively small.   To understand this, we may relate the lensing effects on the three-dimensional brightness distribution to that of the two-dimensional power spectrum.  In the limit of the flat sky\footnote{The corresponding expression accounting for all-sky curvature is given in \citet{zfh04}.}, the two-dimensional power spectrum at a specific redshift ($C^T(\Delta\theta) = \int d^2 l/(2\pi)^2 \, C^T_l e^{i\bm{l}\cdot\bm{\Delta\theta}}$) is related to the three-dimensional spectrum ($\langle T(\bm{r}) T(\bm{r'})\rangle = \int d^3 k /(2\pi)^3 \, P(k) e^{i\bm{k} \cdot \bm{\Delta r}}$) as $C^T_l = \int_0^\infty d k_\parallel/(\pi D^2)\,  P(l/D, k_\parallel)$.
That is, for a projected wavenumber $l$, the two-dimensional spectrum is simply the sum of the power over all $k_\parallel$ for all modes with transverse component $k_\perp = l/D$.  From this a relation between the three-dimensional lensed fraction, $\Delta^P(\bm{k})$, and the two-dimensional lensed fraction, defined as $\Delta^C_l \equiv (\tilde{C}^T_l - C^T_l)/C^T_l$, can be derived:
\begin{equation}
\Delta^C_l = \int_0^\infty d k_\parallel\, \Delta^P(l/D, k_\parallel) \rho(l/D, k_\parallel)
\end{equation}
\begin{equation}
\rho(l/D, k_\parallel) = \frac{P(l/D, k_\parallel)}{ \int_0^\infty d k_\parallel P(l/D, k_\parallel)}
\end{equation}
where $\rho(k_\perp, k_\parallel)$ is a weighting function proportional to the three-dimensional power spectrum and normalized by the total variance in modes with transverse component $k_\perp = l/D$, such that $\int_0^\infty d k_\parallel \rho(l/D, k_\parallel) = 1$.  

From this equation, one can understand how the two-dimensional lensing effects are generated from the three-dimensional lensing distortions.  We have seen how the three-dimensional lensing involves the mixing of the power in different $k_\perp$ modes for each group of modes with the same fixed $k_\parallel$.  The two-dimensional lensing effect is then simply the weighted sum of the lensed fractions of different $k_\parallel$ modes with the same fixed $k_\perp = l/D$.  In Figure \ref{rho} we show the normalized weight function $\rho(k_\perp, k_\parallel)$ for several values of $k_\parallel$.  The shape can be understood in the same way as the effective lensed spectra are understood, but reversing the roles of $k_\perp$ and $k_\parallel$.  On scales $k_\parallel \ll k_\perp$, the weight is $\rho \sim \text{constant}$ with respect to $k_\parallel$, while on scales $k_\parallel \gg k_\perp$, $\rho \sim P(k_\parallel)$, which is a declining function of wavenumber on small scales.  At a given transverse scale $k_\perp = l/D$, the weight function $\rho$ clearly favors contributions from modes $\alpha \gtrsim \pi/4$, and discounts more inclined modes $\alpha \lesssim \pi/4$.  Thus, the magnitude of the two-dimensional lensed fraction of $\tilde{C}^T_l$ is primarly an average of the three-dimensional lensed fractions for the near transverse wavevectors, and is not affected by the highly inclined modes that are more strongly modified through lensing.  Finally, note that the above formalism will be inaccurate at scales $l \lesssim 100$ because of sky curvature effects, however the intuition will be the same.

\subsection{Angular Structure}

\begin{figure}[t]
\centering
\includegraphics[angle=90,scale=0.4]{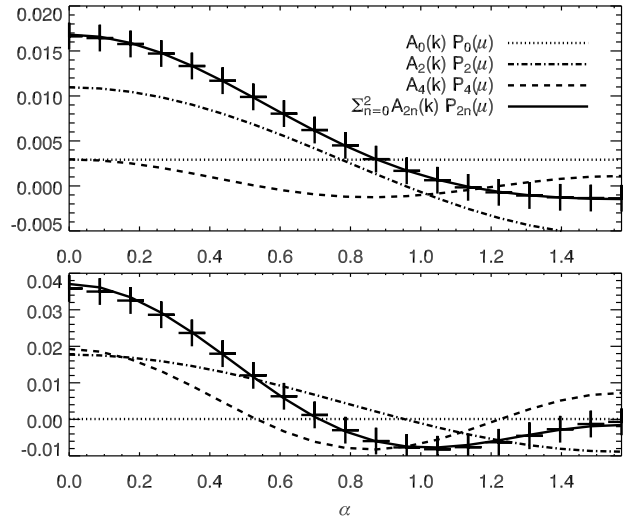}
\caption{\label{angprofile} Angular dependence of the three-dimensional lensed fraction. Here we illustrate the numerical solution (crosses), the three Legendre polynomial components, $P_{2n}(\mu = \cos\alpha)$, and their sum (solid line).  This pertains to the $z =15$ model for $k = 1 \text{ h Mpc}^{-1}$ (top) and $k = 10 \text{ h Mpc}^{-1}$ (bottom). }
\end{figure}

\begin{figure}[t]
\centering
\includegraphics[angle=90,scale=0.3757]{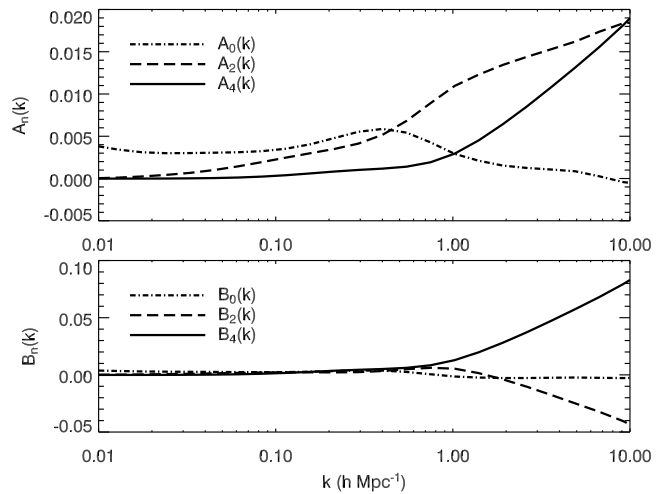}
\caption{\label{coeffs} The angular components of the lensed fraction of the three-dimensional power spectrum.  This pertains to the $z =15$ model.  The coefficents are defined such that the lensed fraction is $\Delta^P(\bm{k}) = A_0(k) + A_2(k) P_2(\mu) + A_4(k) P_4(\mu) = B_0(k) + B_2(k) \mu^2 + B_4(k) \mu^4$, where $\mu = \cos\alpha$.}
\end{figure}

In this section, we examine more closely the angular structure of the three-dimensional lensed fraction.  In Figure \ref{angprofile}, we show the lensed fraction $\Delta^P$ as a function of the orientation angle $\alpha$ for two values of $k$.  There is no obvious analytic dependence on the angle from the solution, Eq. \eqref{pobvious}.  However, there is a fairly smooth transition of the lensed fraction from highly inclined modes to nearly transverse modes.  This suggests a simple expansion in a few Legendre polynomials $P_l(\mu)$ in $\mu = \cos\alpha$.
\begin{equation}
\Delta^P(k,\alpha) = A_0(k) + A_2(k)P_2(\mu) + A_4(k) P_4(\mu)
\end{equation}
Here we have expanded only in even polynomials since the lensed fraction is symmetric around $\alpha = \pi/2$, and the coefficient functions $A_{2n} (k)$ are determined numerically using the completeness relations.  Shown in Figure \ref{angprofile} are the separate components, $A_{2n}$, and their sum.  It is evident that this truncated expansion well describes the angular dependence of the lensed fraction.
This expansion is equivalent to a fourth order polynomial in $\mu$.
\begin{equation}
\Delta^P(k,\alpha) = B_0(k) + B_2(k)\mu^2 + B_4(k)\mu^4
\end{equation}
with $B_0(k) = A_0(k) - (1/2) A_2(k) + (3/8) A_4(k)$, $B_2(k) = (3/2) A_2(k) - (30/8) A_4(k)$, and $B_4(k) = (35/8) A_4(k)$. These coefficients are plotted for an example in Figure \ref{coeffs}.

\citet{barkanaloeb05} have shown that accounting for redshift space distortions due to peculiar velocities leads to angular anisotropic components in the intrinsic power spectrum that can be expressed as a polynomial in $\mu = \cos\alpha$, i.e.
\begin{equation}\label{mudecomp}
P(\bm{k}) = P_{\mu^0}(k) + P_{\mu^2}(k) \mu^2 + P_{\mu^4}(k) \mu^4.
\end{equation}
The measurement of the separate components allows the separation of the various astrophysical effects that contribute to the 21 cm fluctuations.  We have shown that gravitational lensing of the purely isotropic component ($P(k) = P_{\mu^0}(k)$) generates components with the same angular dependences, with magnitudes described by the coefficients $B_n(k)$.  This effect arises from the different effective transverse spectra $k_\perp^2 P(k_\perp, k_\parallel)$ that are lensed for different orientations of $\bm{k}$.  Future measurements of the $\mu^n$ components of the power spectrum will need to take into account the gravitational lensing contributions to each component when precisions better than half a percent are achieveable for scales $k \lesssim 1 \text{ h Mpc}^{-1}$ and when precisions of order one percent are achieveable for scales $k \gtrsim 1 \text{ h Mpc}^{-1}$.  

\subsection{Lensing of Intrinsic Anisotropies}

In the previous sections we have described the structure of weak lensing effects on a spherically symmetric power spectrum.  This approximation will be most relevant late in reionization when fluctuations from reionized bubbles will dominate over angular anistropies caused by peculiar velocity distortions.  Early in reionization, when bubbles are less important, the angular components that vary with $\mu^2$ and $\mu^4$ can be of the same order and larger than the spherically symmetric part.  In this case, in order to describe the gravitational lensing effects, one must consider the full anisotropic intrinsic power spectrum, Eq. \eqref{mudecomp}.  Although on the surface this appears much more complicated, the equations derived in previous sections to calculate the lensing are straightfowardly applicable to any three-dimensional power spectrum with a general $P(k_\perp, k_\parallel)$ dependence.

We have considered lensing of an early 21 cm model, in the case where $P_{\mu^2}(k)$, $P_{\mu^4}(k) \gtrsim P_{\mu^0}(k)$.  We find that the lensed fraction is of the same order of magnitude as those described in the spherically symmetric case (Fig. \ref{3dlensing}).  The overall effect is smaller than one percent on scales $k \lesssim 1 \text{ h Mpc}^{-1}$, but highly inclined modes can be lensed $\gtrsim 1\%$ on scales $k \gtrsim 1 \text{ h Mpc}^{-1}$.  The lensing again generates angular components depending on $\mu^{2n}$, with amplitudes that grow with increasing $k$.  The angular profiles are qualitatively similar to those shown in Figure \ref{angprofile}, however we find that higher order Legendre polynomials, i.e. $P_l(\mu)$ with $l > 4$, are needed to fully describe the angular variation of the lensing at smaller scales. 

These characteristics can be understood qualitatively.  We expect that the order of magnitude of the lensing to remain the same as long as the $P_{\mu^2}(k)$ and $P_{\mu^4}(k)$ components are also smooth and do not add sharp features to the overall spectrum.  
Next, we note that the lensing in the regime of highly inclined modes at large $k_\parallel \gg k_\perp$ is the same as described in the spherically symmetric case.  This is simply because, in this regime, $\mu^2 = \cos^2 \alpha = k_\parallel^2 /(k_\parallel^2 + k_\perp^2) \approx 1$, so that the intrinsic power spectrum is $P(\bm{k}) \approx P_{\mu^0}(k_\parallel) + P_{\mu^2}(k_\parallel) + P_{\mu^4}(k_\parallel)$, since $k = (k_\perp^2 + k_\parallel^2)^{1/2} \approx k_\parallel$.  Thus, the transverse gradient spectrum, $k_\perp^2 P(\bm{k}) \sim k_\perp^2$ is effectively Poisson for $k_\perp \ll k_\parallel$, and hence will be lensed by a positive constant determined by the convergence as before.  This effect becomes most pronounced for very large $k_\parallel$.  

The more complicated structure of the lensing is intuitively a reflection of the more complex intrinsic spectrum.  Just as the lensing of the purely isotropic component generates lensed components varying as $\mu^2$ and $\mu^4$, we may expect that lensing an intrinsic spectrum with significant $\mu^2$ and $\mu^4$ pieces will generate higher order angular moments.  The general structure of the lensed fraction is determined by the shape of the transverse gradient spectrum.  For a highly anisotropic intrinsic spectrum, there are three distinct regimes that determine the overall shape of $k_\perp^2 P(k_\perp, k_\parallel)$ with respect to $k_\perp$, for a given slice of modes with fixed $k_\parallel$.  In the first regime $k_\perp \ll k_\parallel$, and the spectrum is locally Poisson.  In the second regime, $k_\perp \sim k_\parallel$, the factor $\mu^2$ is beginning to decrease but is still a fraction of order unity.  Here, the spectrum is determined by the shape of the $k_\perp^2 (P_{\mu^2}(k)\mu^2 + P_{\mu^4}(k)\mu^4)$ components versus $k_\perp$.  In the third regime, $k_\perp \gg k_\parallel$, so that the $\mu^2$ terms are tiny and suppress the contributions from $P_{\mu^2}(k)$ and $P_{\mu^4}(k)$.  Here, the modes are mostly transverse on the sky, and the transverse gradient spectrum is dominated by the isotropic term, $k_\perp^2 P(\bm{k}) \approx k_\perp^2 P_{\mu^0}(k_\perp)$.  These three regimes and the transitions between them account for the additional structure that can be seen in the lensed fraction.

\section{Detectability}
\label{detectsection}
In this section we discuss the magnitude of the gravitational lensing effects on the three-dimensional power spectra in relation to the expected sensitivites of planned radio array experiments.  We have seen how the lensing effects are typically $\lesssim 1\%$ on scales $k \lesssim 1 \text{ h Mpc}^{-1}$.  The first generation of experiments is unlikely to achieve this precision in the near future, and thus lensing effects will be unimportant.  We consider here the potential sensitivity of a SKA experiment to an EoR signal at $z=8$, with a mean ionized fraction of 0.54, and we only consider the lensing of the spherically symmetric part of the power spectrum $P(\bm{k})$.  The parameters of the SKA are not fixed; here we assume an array of 5000 antennae with a total effective area of $6 \times 10^5 \text{ m}^2$, and a minimum frequency resolution of $\Delta\nu = 0.01 \text{ MHz}$.  The density of antennae is assumed to fall off as $r^{-2}$ from the center.  The total bandwidth of the observation is $6 \text{ MHz}$ and the integration time is $1000$ hours.  The system temperature is set to $T_{\text{sys}} = 440 \text{ K}$ \citep{bowmanmoraleshewitt05} and the effective survey size is $\pi 5.6^2 \text{ deg}^2$.

\begin{figure}[t]
\centering
\includegraphics[scale=0.4]{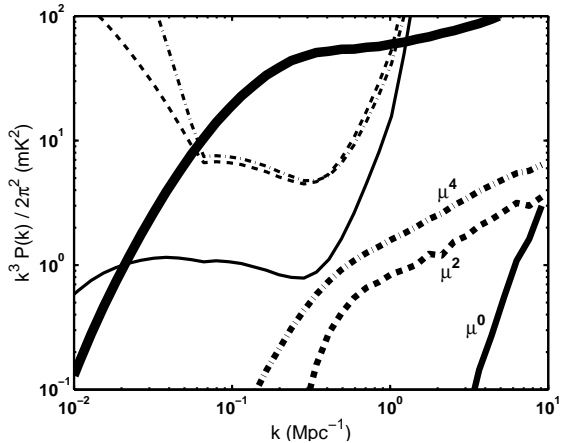}
\caption{\label{detectable} Sensitivity estimates for a 21 cm power spectrum at $z = 8$.  The very thick solid line is the isotropic component of the power spectrum $P(\bm{k})$.  The thick solid, dashed, and dash-dotted curves labelled $\mu^0$, $\mu^2$, and $\mu^4$ correspond to the contributions due to lensing, $B_0(k) P(k)$, $B_2(k) P(k)$, $B_4(k) P(k)$, respectively.  The thin solid, dashed, and dash-dotted curves are the sensitivity estimates for each component respectively.  Here we have assumed an SKA configuration as described in the text.}
\end{figure}

In order to compute the potential sensitivity to the lensed three-dimensional power spectrum $\tilde{P}(\bm{k})$, we follow the methods described in \citet{morales05} and \citet{mcquinn05}.  The contributions from sample variance and thermal noise in visibilities to the uncertainty in the power spectrum $\delta P(k,\alpha)$ in annuli at fixed $k$ and $\alpha$ are computed in logarthmically spaced bins in $k$.  For simplicity we have ignored here the complicated matter of foregrounds.  Next the uncertainty in the components of the model $\tilde{P}(\bm{k}) = \lambda_1(k) + \lambda_2(k) \mu^2 + \lambda_3(k) \mu^4$ is computed.  Here $\lambda_1(k) \equiv (1+B_0(k))P(k)$, $\lambda_2(k) \equiv B_2(k) P(k)$, and $\lambda_3(k) \equiv B_4(k) P(k)$.  The Fisher matrix is
\begin{equation}
F_{ij}(k) = \sum_\alpha \frac{1}{\delta P(k,\alpha)^2} \frac{\partial \tilde{P}}{\partial \lambda_i(k)} \frac{\partial \tilde{P}}{\partial \lambda_j(k)}.
\end{equation}
An estimate of the uncertainty on the components $\lambda_i$ is $F^{-1/2}_{ii}$.
In Figure \ref{detectable} these estimates are plotted.  The component contributions from gravitational lensing are smaller than the expected sensitivities by at least a factor of 10 on scales $k \sim 0.3 \text{ Mpc}^{-1}$.  
If future experiments are to detect the gravitational lensing modifications of the 21 cm power spectrum, they will need greater angular and frequency resolution in order to probe the regime $k \gtrsim 1 \text{ h Mpc}^{-1}$, where lensing effects become relatively larger and lensing contributions to the angular components become more pronounced. If futuristic experiments are able to probe the power spectrum components to $\lesssim 1 \%$ precision, gravitational lensing effects will be important to take into account.

\section{Summary and Conclusion}
\label{conclusion}

In this paper, we have addressed the effects of gravitational lensing of large-scale structure on the brightness temperature power spectrum of high redshift 21 cm.  Gravitational lensing presents a secondary source of fluctuations caused by density perturbations between emission and observation, and reprocesses intrinsic fluctuations caused by density, neutral fraction, and spin-temperature fluctuations in the EoR.

We have presented a new formulation for calculating the gravitational lens effect on temperature power spectra of diffuse backgrounds.  This harmonic-space approach leads to a hierarchical system of equations that describe the lensing effect on the power spectrum as one integrates differential shells of lensing modes regulated by a single cutoff scale $\Lambda$.  We have described the general structure of the system of equations, which can be closed at the chosen level by approximating the magnification matrix.  This method is an alternative to previous series solutions, which require evaluation of high-dimensional integrals or multiple transforms between position space and harmonic space.

We have applied this method to the case of 21 cm fluctuations from the era of reionization.  The effect on two-dimensional spectra $C_l$ at a single frequency (redshift) is typically less than $1 \%$ on scales $100 < l < 10^5$.  In particular we find that higher-order corrections are negligible on all scales despite the dominance of small scale power.  At late times when the spectra are dominated by the ionized bubbles, the flattening of the power spectrum on small scales suppresses the lensing effect due to near scale-invariance of the spectrum.

For small surveys, the three-dimensional power spectrum $P(\bm{k})$ can be directly probed by mapping frequency to line-of-sight distance.  We have shown how the expected lensing effect on the three-dimensional power spectrum can be reduced to a two-dimensional calculation for each set of temperature modes with the same line-of-sight component wavenumber $k_\parallel$.  Since the relevant quantity is the transverse gradient power spectrum, which is different for each $k_\parallel$, gravitational lensing will affect transverse wavevectors differently from line-of-sight wavevectors.  The result is a dependence of the lensed fraction on the angle between $\bm{k}$ and the line-of-sight that is well described by a quartic polynomial in the cosine.  Thus, weak lensing generates angular components of the power spectrum with amplitudes of order one percent or more on small scales, $k \gtrsim 1 \text{ h Mpc}^{-1}$.

Finally, we have compared the predicted lensing signatures to the potential sensitivity limits of SKA observations of the 21 cm emission.  Future experiments will likely need greater sensitivity and resolution in angular and frequency space in order to detect the effects of weak lensing.  The gravitational lensing effects on power spectra will become important for futuristic high-precision 21 cm experiments that can measure the power spectrum to better than one percent and can probe scales $k \gtrsim 1 \text{ h Mpc}^{-1}$.

\acknowledgements

We are grateful to Matt McQuinn for invaluable help with the experiment sensitivity calculation and for providing power spectrum models. 
This work was supported by The David and Lucile Packard
Foundation award 2001-19070A, The Alfred P. Sloan Foundation grant BR-4375, 
NSF grant AST-0506556, and NASA grant NNG05GJ40G.

\bibliographystyle{apj}

\end{document}